\documentclass[prd,aps,preprintnumbers,floats,floatfix,superscriptaddress,preprintnumbers,
showpacs,eqsecnum,nofootinbib,twocolumn
]{revtex4-1}
\usepackage[utf8]{inputenc}
\usepackage{latexsym,array,theorem,mathrsfs,bm,float}
\usepackage{psfrag}
\usepackage{amsfonts,amsmath,amssymb,latexsym,array,afterpage,
theorem,mathrsfs,bm,float,epsfig,color,graphicx,tabularx,here,multirow}

\newcommand{\nn}{\nonumber \\}
\newcommand{\bea}{\begin{eqnarray}}
\newcommand{\ena}{\end{eqnarray}}
\newcommand{\beann}{\begin{eqnarray*}}
\newcommand{\enann}{\end{eqnarray*}}

\newcommand{\RB}{\overset{\scriptscriptstyle  (0)}{R}}

\newcommand{\Rf}{\overset{ \scriptscriptstyle (1)}{R}}
\newcommand{\GB}{\overset{\scriptscriptstyle (0)}{G}}

\newcommand{\gB}{\overset{\scriptscriptstyle  (0)}{g}}

\newcommand{\TB}{\overset{\scriptscriptstyle (0)}{T}}
\newcommand{\cTB}{\overset{\scriptscriptstyle  (0)}{{\cal T}}}

\newcommand{\Rff}{\overset{\scriptscriptstyle (2)}{R}}

\newcommand{\nablaB}{\overset{\scriptscriptstyle  (0)}{\nabla}}

\newcommand{\SB}{\overset{\scriptscriptstyle  (0)}{S}}
\newcommand{\cSB}{\overset{ \scriptscriptstyle (0)}{\cal S}}
\newcommand{\Sf}{\overset{\scriptscriptstyle (1)}{S}}
\newcommand{\cSf}{\overset{\scriptscriptstyle (1)}{\cal S}}
\newcommand{\Sff}{\overset{\scriptscriptstyle (2)}{S}}

\newcommand{\cEff}{\overset{\scriptscriptstyle (2)}{ \mathcal{E} }}

\newcommand{\nablag}{\overset{\scriptscriptstyle  (g)}{\nabla}}
\newcommand{\nablaf}{\overset{\scriptscriptstyle (f)}{\nabla}}

\begin{document}

\baselineskip=12pt
\preprint{WU-AP/1702/17}

\title{Condensate of Massive Graviton and Dark Matter}
\author{Katsuki \sc{Aoki}}
\email{katsuki-a12@gravity.phys.waseda.ac.jp}
\affiliation{
Department of Physics, Waseda University,
Shinjuku, Tokyo 169-8555, Japan
}

\author{Kei-ichi \sc{Maeda}}
\email{maeda@waseda.ac.jp}
\affiliation{
Department of Physics, Waseda University,
Shinjuku, Tokyo 169-8555, Japan
}

\date{\today}

\begin{abstract}
We study coherently oscillating massive gravitons in the ghost-free bigravity theory. This coherent field can be interpreted as a condensate of the massive gravitons. We first define the effective energy-momentum tensor of the coherent massive gravitons in a curved spacetime. We then study the background dynamics of the universe and the cosmic structure formation including the effects of the coherent massive gravitons. We find that the condensate of the massive graviton behaves as a dark matter component of the universe. From the geometrical point of view the condensate is regarded as a spacetime anisotropy. Hence, in our scenario, dark matter is originated from the tiny deformation of the spacetime. We also discuss a production of the spacetime anisotropy and find that the extragalactic magnetic field of a primordial origin can yield a sufficient amount for dark matter.
\end{abstract}



\maketitle

\section{Introduction}
The existence of gravitational waves was indeed confirmed by the direct detections~\cite{Abbott:2016blz,TheLIGOScientific:2016pea}, and their quantum counterpart is called gravitons. The gravitons are defined by perturbations around a background spacetime. The effective energy-momentum tensor of the high-frequency gravitons in General Relativity (GR) was derived by Isaacson~\cite{Isaacson:1967zz,Isaacson:1968zza} which enables us to treat the gravitons as massless spin-2 particles whose energy and momentum change the background geometry. Due to the nonlinear features of the Einstein equations the effective energy-momentum tensor cannot be straightforwardly defined. The gravitons are well-defined when their frequencies (and their momenta) are high enough compared with the curvature scale of the background and then the energy-momentum tensor is defined via a non-local operation which projects the nonlinear quantities of the gravitons onto those in low-frequency modes. However, the low energy states of gravitons, i.e., low frequency/momentum modes of gravitons, should be ill-defined in GR. This is not the case when a graviton is massive. 

Although GR is now widely accepted as a low-energy effective theory of gravity, the question whether the graviton is indeed massless or not has been long discussed (see \cite{Hinterbichler:2011tt,deRham:2014zqa,Schmidt-May:2015vnx} for reviews and \cite{Will:2014kxa,Murata:2014nra,deRham:2016nuf} for experimental constraints on the graviton mass). The linear theory of the massive spin-2 field was constructed by Fierz and Pauli in 1939~\cite{Fierz:1939ix}. Since the gravity must be represented by a nonlinear theory of the metric tensor, the Fierz-Pauli theory requires an extension to the nonlinear theory of the metric in order to obtain the theory of the massive graviton. Generic nonlinear extension of the Fierz-Pauli theory turns to be unstable, called the Boulware-Deser ghost~\cite{Boulware:1973my}. However, the ghost-free nonlinear massive gravity was proposed by de Rham et al. in 2010~\cite{deRham:2010ik,deRham:2010kj} which was further extended into the bigravity theory~\cite{Hassan:2011zd} and the multi-gravity theory~\cite{Hinterbichler:2012cn}. In the bigravity theory or the multi-gravity theory the gravity is still a long-range force because there exists a massless graviton in addition to the massive graviton(s). In the present paper, we focus on the bigravity theory which contains a massless graviton and a massive graviton. The effective energy-momentum tensors of both massless and massive gravitons are defined in the similar way to the case of GR~\cite{Aoki:2016zgp}.

The bigravity theory has received much attentions related to the discovery of dark energy and dark matter. If the graviton mass is extremely small such as $m\sim 10^{-33}$~eV, the present accelerating expansion of the Universe can be explained by the tiny graviton mass~\cite{Volkov:2011an,vonStrauss:2011mq,Berg:2012kn,Comelli:2011zm,Maeda:2013bha,Akrami:2012vf,Akrami:2013ffa,Aoki:2013joa}. Other range of the mass may explain the origin of dark matter. For instance, dark matter is originated from a matter field in the ``dark sector'' when $m\gtrsim 10^{-27}$~eV~\cite{Aoki:2013joa,Aoki:2014cla} whereas the massive graviton itself is a candidate of dark matter when $10^{-4}\, {\rm eV} \lesssim m \lesssim 10^{7}$~eV~\cite{Aoki:2016zgp} (see also~\cite{Babichev:2016hir,Babichev:2016bxi}).

The first suggestion to dark matter in the ghost-free bigravity theory was given by~\cite{Maeda:2013bha} which found that the anisotropy of the spacetime behaves like a dust fluid as for the contribution to the Friedmann equation. However, the following questions have not been cleared: Why does the anisotropy behave as a non-relativistic fluid? Whether or not can it explain other phenomena of dark matter, e.g., the cosmic structure formation? In the present paper, thus, we explore those questions and find that the dark matter component can be regarded as the ``condensate'' of the massive graviton and it can give local structures of the Universe.

We shall focus on the case when the massive graviton is dominated by the zero momentum mode; that is, the configuration of the massive graviton is almost homogeneous. This configuration can be interpreted as the condensate of the massive graviton which we call the massive graviton condensate. Contrary to the case of the massless graviton, the zero momentum mode of the massive graviton shows a coherent oscillation due to the mass term. Therefore, we can define the energy-momentum tensor of the zero momentum mode of the massive graviton as long as the graviton mass is larger than the curvature scale of the background spacetime. We find that the zero momentum mode of the massive graviton gives a dark matter contribution to the Fridemann equation and the tiny fluctuations around the zero momentum mode provide the cosmic structure formation. The constraint on the graviton mass to be dark matter is the same as that obtained in \cite{Aoki:2016zgp}, i.e., $10^{-4}\, {\rm eV} \lesssim m \lesssim 10^{7}$~eV, in general.

From the geometrical aspect the zero momentum mode of the massive graviton represents the anisotropy of the spacetime. The universe filled with the zero momentum massive gravitons is interpreted as a homogeneous spacetime. The anisotropic component of the universe acts as dark matter which is indeed shown by \cite{Maeda:2013bha}. The metric perturbations around the homogeneous spacetime can provide the structures of the universe.

The tiny anisotropy of the universe can be produced when there is a coherent field with an anisotropic stress. A possible candidate of the source is the extragalactic magnetic field of a primordial origin~(see e.g., \cite{Vachaspati:1991nm,Ratra:1991bn,Garretson:1992vt,Sigl:1996dm}). Recent blazar observations implies the existence of the extragalactic magnetic field whose lower bound of the strength $B_0$ is about $10^{-17}$ G \cite{Neronov:1900zz,Tavecchio:2010mk,Dolag:2010ni,Essey:2010nd,Taylor:2011bn,Takahashi:2013lba,Chen:2014rsa}. This magnetic field could be produced in the early universe~\cite{Fujita:2016qab,Adshead:2016iae}. We will show that the coherent magnetic field can yield a sufficient amount of the massive graviton condensate in order to explain the present abundance of dark matter.

The paper is organized as follows. After a brief introduction about the ghost-free bigravity theory in Sec.~\ref{bigravity}, we define the effective energy-momentum tensor of the coherent massive graviton in Sec.~\ref{coherent_TG}. The homogeneous configuration of $\varphi_{\mu\nu}$ is studied in Sec.~\ref{homogeneous_MGC} which reproduces the result obtained by \cite{Maeda:2013bha} from a field theoretical aspect. We then study the perturbations around the homogeneous mode in Sec.~\ref{structure}. We show that the massive graviton condensate is indeed a viable candidate of dark matter. In Sec.~\ref{production}, a production of the condensate to be dark matter is discussed. We give a summary and some discussions in Sec. \ref{summary}. In Appendix \ref{sec_Tmunu}, we summarize the definitions of the energy-momentum tensors of the high-frequency massive and massless gravitons in a curved spacetime. We briefly study the Bianchi I universe in bigravity in Appendix \ref{sec_Bianchi}. In Appendix \ref{sec_general_pert}, we detail the calculations about the inhomogeneous modes of the massive graviton condensate.

\section{Bigravity theory}
\label{bigravity}
The action of the bigravity theory proposed by Hassan and Rosen \cite{Hassan:2011zd} is given by
\begin{eqnarray}
\!\!\!\!\!\!\!\!\!\!  S &=&\frac{1}{2 \kappa _g^2} \int d^4x \sqrt{-g}R(g)+ \frac{1}{2 \kappa _f^2}
 \int d^4x \sqrt{-f} \mathcal{R}(f) \nonumber \\
&-&
\frac{m^2}{ \kappa ^2} \int d^4x \sqrt{-g} \mathscr{U}(g,f) 
+S^{[\text{m}]}\,,
\label{action}
\end{eqnarray}
where $g_{\mu\nu}$ and $f_{\mu\nu}$ are two dynamical metrics, and
$R(g)$ and $\mathcal{R}(f)$ are their Ricci scalars.
The parameters  $\kappa_g^2=8\pi G$ and $\kappa_f^2=8\pi \mathcal{G}$ are 
the corresponding gravitational constants, 
while $\kappa$ is defined by $\kappa^2=\kappa_g^2+\kappa_f^2$.

 The ghost-free interaction term between the two metrics
is given by
\begin{equation}
\mathscr{U}(g,f)=\sum^4_{k=0}b_k\mathscr{U}_k(\sqrt{g^{-1}f})
\,,
\end{equation}
where $\{b_k\}\,(k=0\, \mbox{-}\, 4)$ are coupling constants
and the 4$\times$4 matrix $\sqrt{g^{-1}f}=\left(\sqrt{g^{-1}f}\right)^{\mu}{}_{\nu}$ is 
defined by 
\begin{equation}
\left( \sqrt{g^{-1}f}\right)^{\mu}{}_{\rho} \left(\sqrt{g^{-1}f}\right)^{\rho}{}_{\nu}
=g^{\mu\rho}f_{\rho\nu}
\,, 
\label{gamma2_metric}
\end{equation}
while $\mathscr{U}_k$ are
the elementary symmetric polynomials of the eigenvalues of the matrix
 $\sqrt{g^{-1}f}$.

Just for simplicity, we assume that matter is coupled only to the $g$-metric
\begin{align}
S^{[\text{m}]}
=S_g^{[\text{m}]}(g,\psi_g)
\,.
\end{align}
We shall briefly discuss the case when other types of matter fields are introduced in Sec.~\ref{summary} and Appendix \ref{sec_Tmunu}. Our conclusion is not changed even for those cases.

The fully nonlinear equations of motion are given by
\begin{align}
G^{\mu\nu}(g)&=\kappa_g^2 \left( T_{\rm (int)}^{\mu\nu}+T^{\mu\nu} \right)
\,, \\
\mathcal{G}^{\mu\nu}(f)&=\kappa_f^2 \mathcal{T}_{\rm (int)}^{\mu\nu}
\,,
\end{align}
where $T^{\mu\nu}$ is the matter energy-momentum tensor while $T_{\rm (int)}^{\mu\nu}$ and $\mathcal{T}_{\rm (int)}^{\mu\nu}$ are derived by the variations of the interaction term $\mathscr{U}$ with respect to $g_{\mu\nu}$ and $f_{\mu\nu}$, respectively. The contracted Bianchi identity and the matter conservation law $\nablag{}_{\mu} T^{\mu\nu}=0$ lead to
\begin{align}
\nablag{}_{\mu} T^{ {\rm (int)} \mu\nu}=0
\,,  \quad
\nablaf{}_{\mu} \mathcal{T}^{ {\rm (int)} \mu\nu}=0
\,,
\end{align}
where $\nablag{}_{\mu}$ and $\nablaf{}_{\mu}$ are the covariant derivatives with respect to $g_{\mu\nu}$ and $f_{\mu\nu}$, respectively.

There is a particular vacuum solution in which two spacetimes are homothetic such that
\begin{align}
f_{\mu\nu}=\xi_0^2  g{}_{\mu\nu} \,,
\end{align}
where $\xi_0$ is a root of the quartic equation
\begin{align}
\Lambda_g=\xi_0^2 \Lambda_f\,,
\label{homothetic_condition}
\end{align}
with
\begin{align}
\Lambda_g&:=m^2\frac{\kappa_g^2}{\kappa^2}(b_0+3b_1 \xi_0+3b_2 \xi_0^2 + b_3 \xi_0^3)
\,, \\ 
\Lambda_f&:=m^2\frac{\kappa_f^2}{\kappa^2} (b_4 +3 b_3 \xi_0^{-1} +3 b_2 \xi_0^{-2} +b_1 \xi_0^{-3})\,.
\end{align}
For the homothetic solutions, we obtain
\begin{align}
T_{\rm (int) }^{\mu \nu}=\Lambda_g g^{\mu\nu}
\,, \quad
\mathcal{T}_{\rm (int)}^{\mu\nu}=\Lambda_f f^{\mu\nu}
\,,
\end{align} 
thus, the constants $\Lambda_g$ and $\Lambda_f$ are effective cosmological constants for the $g$-spacetime and the $f$-spacetime, respectively. 
In what follows, we assume 
\begin{align}
\Lambda_g =\Lambda_f=0\,, \label{no_cc}
\end{align}
because we are interested not in dark energy but in dark matter.
The equations for the homothetic spacetime are exactly reduced into those in GR which indicates that the homothetic solution contains only the massless graviton modes. The degrees of freedom of the massive graviton mode do not exist in the homothetic solution.

\section{Energy-momentum tensor of coherent gravitons}
\label{coherent_TG}
In this section, we derive the effective energy-momentum tensor of the coherently oscillating gravitons focusing on the cosmological situation. General discussion about the energy-momentum tensor of gravitons is given in Appendix \ref{sec_Tmunu}. 

As is well known in GR, when we discuss some structure produced by high frequency graviational waves, we have to separate the high frequency modes from smoothed background.
The length or/and time scale associated with the gravitational waves should be sufficiently shorter than the typical scale of the smooth background~\cite{Isaacson:1967zz,Isaacson:1968zza}. Under this setting, the energy-momentum tensor of gravitational waves is defined by the nonlinear terms of the perturbed Einstein equation averaged over a length or/and time scale. We then obtain the propagating equation for the gravitational waves and the Einstein equation for the background including the backreaction from gravitational waves. We shall apply this procedure to the cosmological setting with the coherently oscillating massive gravitons. In the coherent case, we have to take care which we perform a spatial average or a time average.

We consider the homogeneous universe with tiny metric perturbations
\begin{align}
g_{\mu\nu}&=g^{\rm (hom)}_{\mu\nu}(t)+ \delta g^{\rm (inh)}_{\mu\nu}(t,\mathbf{x}) \,, \nn
f_{\mu\nu}&=f^{\rm (hom)}_{\mu\nu}(t)+ \delta f^{\rm (inh)}_{\mu\nu}(t,\mathbf{x}) \,,  \label{Bianchi+pert}
\end{align}
where $g^{\rm (hom)}_{\mu\nu},f^{\rm (hom)}_{\mu\nu}$ are the metrics of the homogeneous spacetime and $\delta g^{\rm (inh)}_{\mu\nu},\delta f^{\rm (inh)}_{\mu\nu}$ represent the inhomogeneous perturbations. 
Since we are interested in the coherent gravitons, the time coordinate has to be appropriately chosen in order that the $t=$ constant hypersurfaces are given by almost homogeneous spaces. Then, on each hypersurface, the homogeneous parts can be obtained by
\begin{align}
g^{\rm (hom)}_{\mu\nu}=\langle g_{\mu\nu} \rangle_V \,, \quad f^{\rm (hom)}_{\mu\nu}=\langle f_{\mu\nu} \rangle_V \,, \label{homogeneous_parts}
\end{align}
where $\langle \cdots \rangle_V$ is the spatial average where the averaged length scale is assumed to be much lager than the scale of the inhomogeneities.
The dynamics of the homogeneous spacetime in bigravity was studied in \cite{Maeda:2013bha}. Up to the linear perturbation theory, one may directly analyze the dynamics of the perturbations under the ansatz \eqref{Bianchi+pert}. In the present paper, however, we consider another separation of the metrics rather than \eqref{Bianchi+pert}. We first summarize the strategy of our calculations and the explicit analysis are given in Sec.~\ref{homogeneous_MGC} and Sec.~\ref{structure}.

The bigravity theory contains two types of dynamical degrees of freedom, the massless graviton and the massive graviton. First, we separate the metrics $g_{\mu\nu}$ and $f_{\mu\nu}$ into the massless mode and the massive mode. Up to the linear perturbations around the homothetic background, we can introduce the mass eigenstates of the gravitons. However, the definitions of the massless mode and the massive mode of the metrics would be ambiguous in the nonlinear orders in which the gravitons are no longer diagonalized. (see discussions in \cite{Hassan:2012wr,Babichev:2016bxi}). Nevertheless, the massless mode and the massive mode are still meaningful if the perturbative expansion is viable. Therefore, we only consider the situation that the spacetimes are well approximated by the homothetic solution.

We focus on the late stage of the universe such that
\begin{align}
m^2 \gg H^2 \,, \label{Hubble}
\end{align}
where $H$ is the Hubble expansion rate in which the massive graviton has too heavy mass to be excited. Hence, the amplitude of the massive graviton is suppressed and then the metrics $g_{\mu\nu}$ and $f_{\mu\nu}$ are approximated by the homothetic solution (We recall that the homothetic solution give a spacetime without the excitation of the massive graviton). 
We then perturbatively treat the massive mode $g^{\rm (massive)}_{\mu\nu}$ which is defined by the difference between two metrics
\begin{align}
g^{\rm (massive)}_{\mu\nu}=\frac{\alpha}{1+\alpha}\left( g_{\mu\nu}-\xi_0^{-2}f_{\mu\nu} \right)\,,
\end{align}
where $\alpha:=\xi_0^2 \kappa_g^2/\kappa_f^2$ and we assume $|g^{\rm (massive)}_{\mu\nu}| \ll 1$.
On the other hand, the massless mode is given by
\begin{align}
g^{\rm (massless)}_{\mu\nu}=\frac{1}{1+\alpha}\left(g_{\mu\nu}+\alpha \xi_0^{-2} f_{\mu\nu} \right)
\,.
\end{align}
As a result, the metrics $g_{\mu\nu}$ and $f_{\mu\nu}$ can be decomposed into the massless mode and the massive mode as follows:
\begin{align}
g_{\mu\nu}&=g_{\mu\nu}^{\rm (massless)}+g_{\mu\nu}^{\rm (massive)} \,, \nn
f_{\mu\nu}&=\xi_0^2\left( g_{\mu\nu}^{\rm (massless)}-\alpha^{-1} g_{\mu\nu}^{\rm (massive)} \right)\,. \label{mass_sepa} 
\end{align}

We further decompose the massless and the massive modes into the low-frequency modes and the high-frequency modes, respectively:
\begin{align}
g_{\mu\nu}^{\rm (massless)}&=\gB{}_{\mu\nu}+\frac{h_{\mu\nu}}{M_{\rm pl}} \,, \nn
g_{\mu\nu}^{\rm (massive)}&=M_{\mu\nu}+\frac{\varphi_{\mu\nu}}{M_G} \,, \label{low+high}
\end{align}
where the high-frequency modes $h_{\mu\nu}$ and $\varphi_{\mu\nu}$ are normalized by two mass scales
\begin{align}
M_{\rm pl}&:=\frac{\xi_0 \bar{\kappa}}{\kappa_g\kappa_f}\,,\quad
M_G:=\frac{\bar{\kappa}}{\kappa_g^2}=\frac{M_{\rm pl}}{\alpha^{1/2}}\,, 
\end{align}
with $\bar{\kappa}^2=\kappa_g^2+\xi_0^{-2} \kappa_f^2 $.
The low-frequency modes are defined by
\begin{align}
\gB{}_{\mu\nu}=\langle g_{\mu\nu}^{\rm (massless)} \rangle_T \,, \quad M_{\mu\nu}=\langle g_{\mu\nu}^{\rm (massive)} \rangle_T
\end{align}
where $\langle \cdots \rangle_T$ is the time average\footnote{Alternatively, the low-frequency projection operator can be the oscillation average, i.e., the time average over one coherent oscillation $T=2\pi/m$.} over some time interval $T$ which is assumed to be
\begin{align}
m^{-1} \ll T  \ll H^{-1} \,.
\end{align}
Then, the metric tensors $g_{\mu\nu}$ and $f_{\mu\nu}$ are divided into four components: $\gB{}_{\mu\nu},M_{\mu\nu},h_{\mu\nu}$ and $\varphi_{\mu\nu}$. The meaning of each variables is summarized in Table \ref{table_metric}.

\begin{table*}[tb]
\caption{The separations of the metric tensors. }
\label{table_metric}
\begin{tabular}{ccccc}
\hline
            & Low-frequency & High-frequency \\
\hline\hline
Massless mode $g^{\rm (massless)}_{\mu\nu}\,$  & $\gB{}_{\mu\nu}=\bar{g}_{\mu\nu}+\delta g_{\mu\nu}$ & $h_{\mu\nu}/M_{\rm pl}$ \\
Massive mode  $g^{\rm (massive)}_{\mu\nu}\,$ & $M_{\mu\nu}=\bar{M}_{\mu\nu}+\delta M_{\mu\nu} \,\,$ & $\,\,\varphi_{\mu\nu}/M_G=\bar{\varphi}_{\mu\nu}/M_G+\delta \varphi_{\mu\nu}/M_G$ \\
\hline
Homogeneous mode $g^{\rm (hom)}_{\mu\nu}\,$ & $\bar{g}_{\mu\nu}+\bar{M}_{\mu\nu}$ & $\bar{\varphi}_{\mu\nu}/M_G$ \\
Inhomogeneous mode $\delta g^{\rm (inh)}_{\mu\nu}\,$ & $\delta g_{\mu\nu} +\delta M_{\mu\nu} $ & $h_{\mu\nu}/M_{\rm pl}+ \delta \varphi_{\mu\nu}/M_G$ \\
\hline
\end{tabular}
\end{table*}

We briefly mention the relation between two separations \eqref{Bianchi+pert} and \eqref{low+high}. The variables $\gB{}_{\mu\nu},M_{\mu\nu},h_{\mu\nu}$ and $\varphi_{\mu\nu}$ are divided into the homogeneous parts and the inhomogeneous parts
\begin{align}
\gB{}_{\mu\nu}&=\bar{g}_{\mu\nu}(t)+\delta g_{\mu\nu}(t,\mathbf{x}) \,, \nn
M_{\mu\nu}&=\bar{M}_{\mu\nu}(t)+\delta M_{\mu\nu}(t,\mathbf{x}) \,, \nn
h_{\mu\nu}&=h_{\mu\nu}(t,\mathbf{x}) \,, \nn
\varphi_{\mu\nu}&=\bar{\varphi}_{\mu\nu}(t)+\delta \varphi_{\mu\nu}(t,\mathbf{x}) \,, \label{gB}
\end{align}
where the homogeneous parts are defined via the spatial average $\langle \cdots \rangle_V$ as with \eqref{homogeneous_parts}. We then obtain
\begin{align}
g^{\rm (hom)}_{\mu\nu}&=\bar{g}_{\mu\nu}+\bar{M}_{\mu\nu}+\frac{\bar{\varphi}_{\mu\nu}}{M_G} \,, \\
\delta g^{\rm (inh)}_{\mu\nu}&=\delta g_{\mu\nu}+\delta M_{\mu\nu}+\frac{h_{\mu\nu}}{M_{\rm pl}}+\frac{\delta \varphi_{\mu\nu}}{M_G}\,.
\end{align}
It is worth noting that 
\begin{align}
\langle h_{\mu\nu} \rangle_V=0 \,, \quad \langle \varphi_{\mu\nu} \rangle_V =\bar{\varphi}_{\mu\nu}\neq 0\,,
\end{align}
because $h_{\mu\nu}$ is massless while $\varphi_{\mu\nu}$ is massive. The zero momentum mode of the massless graviton cannot be high-frequency whereas that of the massive graviton can be high-frequency due to the coherent oscillation. Since we have assumed that the configuration of the fields are almost homogeneous, the massive graviton is dominated by the zero momentum mode $\bar{\varphi}_{\mu\nu}$, that is,
\begin{align}
|\bar{\varphi}_{\mu\nu}| \gg |\delta \varphi_{\mu\nu}| \,.
\end{align}
We call this configuration of $\varphi_{\mu\nu}$ the massive graviton condensate because a large fraction of $\varphi_{\mu\nu}$ occupies the single zero momentum state $\bar{\varphi}_{\mu\nu}$.

In the separation \eqref{Bianchi+pert}, the ``backgrounds'', i.e., the homogeneous modes $g^{\rm (hom)}_{\mu\nu}$ and $f^{\rm (hom)}_{\mu\nu}$, are obtained by the spatial average whereas the ``background'' in \eqref{low+high}, i.e., the low-frequency massless mode $\gB{}_{\mu\nu}$, is given by the time average and then it can be inhomogeneous. An advantage of the separation \eqref{low+high} is that the high-frequency ``perturbations'' $h_{\mu\nu}$ and $\varphi_{\mu\nu}$ can be treated as tensor fields, propagating on the low-frequency ``background'' $\gB{}_{\mu\nu}$, with well-defined energy-momentum tensors.

The amplitude of the massive graviton is small so we have the inequalities
\begin{align}
|M_{\mu\nu} | \,, \,\, |\varphi_{\mu\nu}|/M_G \ll |\gB{}_{\mu\nu}|\,. \label{small_MG}
\end{align}
The amplitude of $h_{\mu\nu}/M_{\rm pl}$ is also small since $h_{\mu\nu}$ is a part of the inhomogeneity. As a result, we have three small quantities $M_{\mu\nu}, h_{\mu\nu}$ and $\varphi_{\mu\nu}$ which can be treated as the tensors with respect to the ``background'' metric $\gB{}_{\mu\nu}$. We adopt the notation such that the suffices on $M_{\mu\nu}, h_{\mu\nu}$ and $\varphi_{\mu\nu}$ are raised and lowered by $\gB{}_{\mu\nu}$. However, the inequality \eqref{small_MG} does not suggest that the backreaction of $\varphi_{\mu\nu}$ to $\gB{}_{\mu\nu}$ is also small. The orders of magnitude of the Einstein tensor of $\gB{}_{\mu\nu}$ and the energy-momentum tensor of $\varphi_{\mu\nu}$, which we denote $\GB{}_{\mu\nu}$ and $T^{\mu\nu}_G$, are estimated as
\begin{align}
|\GB{}_{\mu\nu} | \sim  H^2  \,, \quad  |T^{\mu\nu}_G| \sim m^2 |\varphi_{\mu\nu}^2|\,,
\end{align}
where $T^{\mu\nu}_G$ is explicitly defined by \eqref{def_TG} below. Thus, if $|\varphi_{\mu\nu}|/M_G \sim  H/m \ll 1$, the massive graviton $\varphi_{\mu\nu}$ can be a dominant component of the universe. In what follows, we assume the massive graviton is the dominant component.

Just for simplicity, we consider the case
\begin{align}
h_{\mu\nu}=0\,.
\label{h=zero}
\end{align}
This is a specific case, but this assumption is reasonable for our interest since the massless gravitons, i.e., the gravitational waves, are sub-dominant in the Universe. To discuss the dynamics of the Universe, the effect of the massless gravitons can be ignored.

To discuss dynamics of the massive graviton condensate $\varphi_{\mu\nu}$, it is sufficient to include the leading and subleading contributions associated with the adiabatic expansion in terms of $m^{-1}$.\footnote{More precisely, we use the dimensionless parameter $H/m$ for the adiabatic expansion. We just refer to $m^{-1}$ as the order of the expansion.} Up to subleading order we can ignore $M_{\mu\nu}$ since the amplitude of $M_{\mu\nu}$ is suppressed by $m^{-2}$ which is the sub-subleading order (see Appendix \ref{sec_Bianchi}). The low-frequency massive mode $M_{\mu\nu}$ gives only negligible contributions.

Ignoring $h_{\mu\nu}$ and $M_{\mu\nu}$, the $g$-spacetime metric is given by
\begin{align}
g_{\mu\nu}= \gB{}_{\mu\nu}+ \frac{\varphi_{\mu\nu}}{M_G}\,. 
\label{physical_metric}
\end{align}
The equations for $\gB{}_{\mu\nu}$ is given by the time-averaged Einstein equation 
\begin{align}
\GB{}^{\mu\nu}=\frac{1}{M_{\rm pl}^2} \left( \TB{}^{\mu\nu}+\langle T^{\mu\nu}_G \rangle_T \right)\,, \label{Einstein_eq}
\end{align}
where the ``effective'' energy-momentum tensors of the matter $\TB{}^{\mu\nu}$ and that of the massive gravitons $T_G^{\mu\nu}$ are defined by the relation
\begin{align}
\left\langle T_{\mu\nu}-\frac{1}{2}g_{\mu\nu} T \right\rangle_T = \TB{}_{\mu\nu}-\frac{1}{2} \gB{}_{\mu\nu} \TB{}_{\alpha \beta} \gB{}^{\alpha\beta}\,, 
\end{align}
and
\begin{align}
T_{G}^{\mu\nu}&= -\left( \gB{}^{\mu\alpha} \gB{}^{\nu\beta}-\frac{1}{2}\gB{}^{\mu\nu} \gB{}^{\alpha\beta} \right) 
 \delta \Rff{}_{\alpha\beta}[\varphi] 
 \nn
&-\frac{m_{\rm eff}^2}{8} \left( 4 \varphi^{\mu\alpha} \varphi^{\nu}{}_{\alpha} - \gB{}^{\mu\nu} \varphi^{\alpha\beta} \varphi_{\alpha\beta} \right) +\mathcal{O}(\varphi^3)\,.
\label{def_TG}
\end{align}
The equation of motion for $\varphi_{\mu\nu}$ is
\begin{widetext}
\begin{align}
\delta \Rf{}_{\mu\nu}[\varphi] 
+\frac{m_{\rm eff}^2}{4}(2\varphi _{\mu\nu}+\varphi^{\alpha}{}_{\alpha} \gB{}_{\mu\nu})+\langle \delta \cEff_{\mu\nu} \rangle_{\rm high} +\mathcal{O}(\varphi^4)
&=\frac{1}{M_G}\left\langle T_{\mu\nu}-\frac{1}{2}g_{\mu\nu} T \right\rangle_{\rm high}\,,
\label{massive_mode}
\end{align}
where the effective graviton mass $m_{\rm eff}$ is defined by
\begin{align}
m_{\rm eff}^2:=m^2 \frac{\bar{\kappa}^2}{\kappa^2}(b_1 \xi_0+2b_2 \xi_0^2 +b_3 \xi_0^3 )\,,
\end{align}
and $\delta \cEff_{\mu\nu}$ include the terms of quadratic in $\varphi_{\mu\nu}$ which is explicitly given by
\begin{align}
\delta \cEff_{\mu\nu} =\frac{\alpha-1}{\alpha^{1/2}M_{\rm pl}}\delta \Rff{}_{\mu\nu}[\varphi]
+\frac{m_{\rm eff}^2}{16\alpha^{1/2}M_{\rm pl}} 
\Big[ &3(1-\alpha)\gB{}_{\mu\nu} \varphi_{\alpha\beta}\varphi^{\alpha\beta}
+4 \{ (1-\beta_2)\alpha-\beta_2  \}  \varphi_{\mu\nu} \varphi^{\alpha}{}_{\alpha}
\nn
&+2\{ (1+2\beta_2)\alpha-(3-2\beta_2)\alpha \} \varphi_{\mu}{}^{\alpha} \varphi_{\nu\alpha} \Big]\,,
\end{align}
\end{widetext}
with
\begin{align}
\beta_2=\frac{b_2\xi_0^2+b_3 \xi_0^3}{b_1 \xi_0 +2b_2 \xi_0^2+ b_2 \xi_0^3} \,.
\end{align}
The symbol $\langle \cdots \rangle_{\rm high}$ denotes a high-frequency projection operator which is given by
\begin{align}
\langle X \rangle_{\rm high}=X-\langle X \rangle_T\,,
\end{align}
for a quantity $X$.
The functionals $\delta \Rf{}_{\mu\nu}$ and $\delta \Rff{}_{\mu\nu}$ are the first order and the second order of the perturbed Ricci curvatures which are explicitly shown in Appendix \ref{sec_Tmunu}.

The amplitude of the coherent oscillation decreases due to the Hubble friction which finally cause the decreasing of the energy density of the massive graviton condensate. To solve \eqref{massive_mode}, we have to retain terms of linear in first derivatives of the metric $\gB{}_{\mu\nu}$.\footnote{When $\varphi_{\mu\nu}$ is treated as a particle, the graviton may be treated as a freely propagating on the flat background since the particle do not feel the effect of the curvature in a small scale. The particle dark matter scenario in bigraivty has been discussed in \cite{Aoki:2016zgp,Babichev:2016hir,Babichev:2016bxi}. } On the other hand, we may ignore terms of higher orders of derivatives of $\gB{}_{\mu\nu}$ which are sub-subleading order contributions; thus, the covariant derivatives commute
\begin{align}
\nablaB{}_{[\alpha} \nablaB{}_{\beta]} \varphi_{\mu\nu}  &\approx 0 \,, \label{commute}
\end{align}
where $\nablaB{}_{\mu}$ is the covariant derivative with respect to $\gB{}_{\mu\nu}$.

Note that the quadratic terms $\delta \cEff{}_{\mu\nu}$ cannot be ignored.
For the homogeneous ansatz, the Friedmann equation schematically reads
\begin{align}
H^2 \sim \frac{1}{M_{\rm pl}^2} m^2 \varphi^2 +\mathcal{O}(\varphi^3)\,,
\end{align}
when the massive graviton is the dominant component of the universe (see Sec.~\ref{homogeneous_MGC} for the explicit expressions). The quadratic term in \eqref{massive_mode} is then
\begin{align}
\delta \cEff{}_{\mu\nu} \sim \frac{m^2}{M_{\rm pl}} \varphi^2 \sim H m \varphi \,,
\end{align}
which yields a comparable effect to the first derivative of the metric. Therefore, we should solve the nonlinear differential equation \eqref{massive_mode} to discuss the dynamics of the coherent massive graviton, in general. 
However, we assume the $Z_2$ symmetry for the self-interactions of the massive graviton: the interaction terms are invariant under the $Z_2$ transformation $\varphi_{\mu\nu}\rightarrow -\varphi_{\mu\nu}$, it prohibits appearance of $\delta \cEff{}_{\mu\nu}$ and then the basic equations become much simpler.
The $Z_2$ symmetry is realized by supposing the symmetry of the gravitational action under the replacement
\begin{align}
g_{\mu\nu} \leftrightarrow f_{\mu\nu} \,, \label{symmetric_gf}
\end{align}
which is realized when
\begin{align}
\kappa_g=\kappa_f \,, \quad b_i=b_{4-i} \,, \;(i=0-4)\,. \label{symmetric_condition}
\end{align}
In this case, $\xi_0=1$ is always a solution to the equation \eqref{homothetic_condition}. For the branch $\xi_0=1$, clearly from the definition of the massive mode, the symmetry \eqref{symmetric_gf} realizes the $Z_2$ symmetry of the massive graviton. Indeed, the parameters \eqref{symmetric_condition} yields $\alpha=1,\beta_2=1/2$ and then
\begin{align}
\delta \cEff{}_{\mu\nu} \equiv 0 \,.
\end{align}
As a result, the equation for the massive mode is linear since the cubic terms can be ignored for our calculations\footnote{The parameters $\alpha$ and $\beta_2$ do not appear at linear order except for the right-hand side of Eq.~\eqref{massive_mode} (we note $M_G=\alpha^{-1/2}M_{\rm pl}$). As a result, the existence of the $Z_2$ symmetry \eqref{symmetric_condition} does not change the theory up to the linear order except for the coupling strength to the matter.}.

By using the normalization of the mass parameter $m$, we can always set
\begin{align}
b_1+2b_2+b_3=1\,,
\end{align}
in which we obtain
\begin{align}
m_{\rm eff}=m
\,,
\end{align}
thus, the mass parameter $m$ indeed corresponds to the graviton mass in the branch $\xi_0=1$. We shall use this normalization in what follows. Combining this normalization with \eqref{no_cc} and \eqref{symmetric_condition}, the coupling constants $b_i$ are expressed by only $b_2$ as
\begin{align}
b_0=b_4=b_2-2\,,\quad
b_1=b_3=\frac{1}{2}-b_2
\,. \label{coupling}
\end{align}

The effective energy-momentum tensor $\TB{}_{\mu\nu}$ is obtained from the smoothing of the true energy-momentum tensor $T_{\mu\nu}$. Even if we assume the true energy-momentum tensor is conserved, i.e., $\nablag_{\mu}T^{\mu\nu}=0$, the smoothed energy-momentum tensor is not conserved, in general, since the energy of the matter can be converted to the one of the graviton and vice versa via the equation \eqref{massive_mode}. The contracted Bianchi identity of \eqref{Einstein_eq} reads the smoothed total energy-momentum tensor is conserved:
\begin{align}
\nablaB{}_{\mu}\left( \TB{}^{\mu\nu}+\langle T^{\mu\nu}_G \rangle_T \right)=0\,.
\end{align}
However, in the late stage of the universe, the massive gravitons must be decoupled from the matter due to the weakness of the gravitational interaction, i.e.,
\begin{align}
\frac{1}{M_G}\left\langle T_{\mu\nu}-\frac{1}{2}g_{\mu\nu} T \right\rangle_{\rm high}
&\approx 0 \,.
\label{free_propagate}
\end{align}
Then, the energy-momentum tensors are individually conserved:
\begin{align}
\nablaB{}_{\mu} \TB{}^{\mu\nu}&\approx 0\,, \\
\nablaB{}_{\mu} \langle T^{\mu\nu}_G \rangle_T &\approx 0\,.
\label{conservation_law}
\end{align}

The conservation of $T^{\mu\nu}_G$ is directly confirmed by using the equation of motion.
For the freely propagating gravitons \eqref{free_propagate}, the equation \eqref{massive_mode} is reduced into
\begin{align}
\left( \nablaB{}_{\alpha} \nablaB{}^{\alpha} -m_{\rm eff}^2 \right) \varphi_{\mu\nu}\approx 0\,, \label{wave_eq} \\
\nablaB{}_{\mu}\varphi^{\mu\nu}\approx 0 \,, \quad \varphi^{\alpha}{}_{\alpha}\approx 0\,. \label{TT_constraint}
\end{align}
Using these equations, one can find
\begin{align}
\nablaB{}_{\mu} T^{\mu\nu}_G \approx 0\,,
\label{div_free}
\end{align}
which is a sufficient condition on the conservation of the graviton energy-momentum tensor \eqref{conservation_law}. We notice, however, that two conservations \eqref{conservation_law} and \eqref{div_free} are not equivalent since \eqref{conservation_law} reads that the smoothed quantity of $T^{\mu\nu}_G$ is conserved. Eq.~\eqref{conservation_law} has information only about macroscopic behavior of $\varphi_{\mu\nu}$ while Eq.~\eqref{div_free} (or \eqref{wave_eq} and \eqref{TT_constraint}) involves information about microscopic behavior.

In the following sections, we will show that there exists a solution such that $\varphi_{\mu\nu}$ behaves as dark matter which explains not only the background dynamics of the universe but also the structure formation. Since we have assumed the inhomogeneities are smaller than the homogeneous modes, we shall linearize the expressions in terms of the inhomogeneities.
For instance, the graviton energy-momentum tensor is expressed as
\begin{align}
T_G^{\mu\nu}=\bar{T}_G^{\mu\nu}(t)+\delta T_G^{\mu\nu}(t, \mathbf{x})\,,
\end{align}
with, in order of magnitude,
\begin{align}
|\delta T_G^{\mu\nu}(t, \mathbf{x})| \sim |  \bar{\varphi}_{\alpha\beta} \partial^2 \delta \varphi_{\mu\nu}  |\,.
\end{align}



\section{Massive graviton condensate as dark matter}
\label{homogeneous_MGC}
In this section, we consider the homogeneous mode $\bar{g}{}_{\mu\nu}$ and $\bar{\varphi}_{\mu\nu}$.
We assume the flat Friedmann-Lema$\hat{\i}$tre-Robertson-Walker (FLRW) background
\begin{align}
\bar{g}_{\mu\nu}dx^{\mu}dx^{\nu}=-dt^2+a^2[dx^2+dy^2+dz^2]\,,
\label{FLRW_ansatz}
\end{align}
and a simple ansatz for $\bar{\varphi}_{\mu\nu}$
\begin{align}
\bar{\varphi}_{\mu\nu}={\rm diag}[0,4a^2\bar{\varphi},-2a^2 \bar{\varphi}, -2a^2 \bar{\varphi} ]
\,, \label{coherent_phi}
\end{align}
where $a$ and $\bar{\varphi}$ are functions of $t$. Note that the ansatz \eqref{coherent_phi} trivially satisfies the constraints \eqref{TT_constraint}.

The massive graviton originally appears from the metric perturbations. The present set up \eqref{coherent_phi} corresponds to considering the axisymmetric Bianchi type I universe in bigravity which we will detail in Appendix \ref{sec_Bianchi} (see \cite{Maeda:2013bha} for more details). 
One may worry about that $\bar{g}{}_{\mu\nu}$ should be also given by the the axisymmetric Bianchi type I universe rather than FLRW universe \eqref{FLRW_ansatz}. However, as we will see just below, the averaged graviton energy-momentum tensor is indeed isotropic and \eqref{coherent_phi} is consistent with \eqref{FLRW_ansatz} (see also \cite{Cembranos:2013cba}). Furthermore, even if one replaces \eqref{FLRW_ansatz} with the Bianchi type universe, its anisotropy decreases as $a^{-6}$ and then the anisotropic part in $\bar{g}{}_{\mu\nu}$ is quickly ignored.

The equation \eqref{wave_eq} including up to the first derivatives of the metric reads
\begin{align}
\ddot{\bar{\varphi}}+3H\dot{\bar{\varphi}}+m^2 \bar{\varphi}= 0\,,
\label{homogeneous_eq}
\end{align}
where $H=\dot{a}/a$. By using \eqref{homogeneous_eq}, we find
\begin{align}
\bar{T}^{tt}_G&=3(\dot{\bar{\varphi}}^2+m^2 \bar{\varphi}^2) \,, \nn
\bar{T}^{xx}_G&=a^{-2}(m^2 \bar{\varphi}^2-\dot{\bar{\varphi}}^2) \,, \nn
\bar{T}^{yy}_G&=\bar{T}^{zz}_G=7a^{-2}(m^2 \bar{\varphi}^2-\dot{\bar{\varphi}}^2) \,, 
\end{align}
and other components are zero.
The approximative solution to \eqref{homogeneous_eq} is 
\begin{align}
\bar{\varphi} = \frac{\bar{\varphi}_0}{a^{3/2}} \cos [m t +\theta_0] 
\end{align}
where $\bar{\varphi}_0$ and $\theta_0$ are integration constants. Since the initial phase $\theta_0$ is not important for the discussion, we set $\theta_0=0$.
After averaging over the time interval $T\gg m^{-1}$, the graviton energy-momentum tensor is calculated as
\begin{align}
\langle \bar{T}_G^{\mu\nu} \rangle_T= {\rm diag}[\bar{\rho}_G ,0,0,0]
\end{align}
where the energy of the massive graviton condensate is
\begin{align}
\bar{\rho}_G = \frac{3}{a^3}m^2 \bar{\varphi}_0^2 \label{rho_G_BG}
\end{align}
Therefore, the massive graviton condensate behaves as a dust fluid. This guarantees the ansatz \eqref{FLRW_ansatz}.

To explain the abundance of dark matter, the amplitude of $\bar{\varphi}$ is required to be
\begin{align}
\bar{\varphi} \sim M_{\rm pl} \frac{H_0}{m} \,,
\end{align}
where $H_0$ is the present Hubble parameter. Since the physical spacetime is given by \eqref{physical_metric}, the universe is filled with the coherent ``gravitational waves'' $\varphi_{\mu\nu}/M_{\rm pl}$ whose dimensionless amplitude and the frequency are 
\begin{align}
| \varphi_{\mu\nu}/M_{\rm pl}| & \sim 10^{-29} \left( \frac{10^{-4}~{\rm eV}}{m} \right)
\,, \\
f &\sim 10^{11} \left(\frac{m}{10^{-4}~{\rm eV}} \right) {\rm Hz}
\,.
\end{align}
The oscillations have too small amplitude and too high frequency and thus there should be no constraint on the existence of $\varphi_{\mu\nu}$ at present.

\section{Cosmic structure formation}
\label{structure}

To study the cosmic structure formation, we then introduce small inhomogeneity to the metric and the massive graviton $\varphi_{\mu\nu}$. Note that the low-frequency ``background'' $\gB{}_{\mu\nu}$ is not necessary to be homogeneous (see \eqref{gB}). 
We treat $\gB{}_{\mu\nu}$ including the inhomogeneity $\delta g_{\mu\nu}$ as the low-frequency massless mode which is verified as long as the momentum of the inhomogeneity is smaller than the graviton mass:
\begin{align}
\frac{k^2}{a^2} \ll m^2\,, 
\end{align}
where $k$ is the comoving momentum of the inhomogeneity defined by \eqref{def_k} later.

For the calculations, we use the adiabatic expansion in terms of the graviton mass inverse $m^{-1}$ (see Section 2.5 in \cite{Gorbunov:2011zzc} for calculations in the case of the scalar condensate and \cite{Cembranos:2016ugq} for the vector case).
We set the orders of both $\bar{g}{}_{\mu\nu}(t)$ and $\delta g_{\mu\nu}(t,\mathbf{x})$ as $\mathcal{O}(m^0)$. Since the time derivatives acting on the low-frequency modes do not change the order of magnitude $m^{-1}$, i.e., $\partial/\partial t =\mathcal{O}(m^0)$, the Friedmann equation leads to $\bar{\rho}_G =\mathcal{O}(m^{0})$. Hence, the homogeneous mode of the massive graviton $\bar{\varphi}_{\mu\nu}$ is of order $\mathcal{O}(m^{-1})$. On the other hand, the amplitude of the inhomogeneous mode of the massive graviton $\delta \varphi_{\mu\nu}$ is of order $\mathcal{O}(m^0)$ as we will show later.

To evaluate the inhomogeneous parts of $T^{\mu\nu}_G$ the coherent background including the sub-leading order is required which is given by
\begin{align}
\bar{\varphi}=\bar{\varphi}_1 \cos[m t]+\bar{\varphi}_2 \sin [mt ] +\mathcal{O}(m^{-3})\,,
\end{align}
where $\bar{\varphi}_1,\bar{\varphi}_2$ are slowly varying functions with
\begin{align}
\bar{\varphi}_1= \frac{\bar{\varphi}_0}{a^{3/2}} = \mathcal{O}(m^{-1}) \,, \quad \bar{\varphi}_2 = \mathcal{O}(m^{-2})\,,
\end{align}
To determine the explicit form of $\bar{\varphi}_2$ we need to solve the equation \eqref{homogeneous_eq} up to the order $\mathcal{O}(m^{-1})$; however, the equation \eqref{homogeneous_eq} is valid up to the order $\mathcal{O}(m^{0})$. Nevertheless, the explicit form of $\bar{\varphi}_2$ is not necessary to evaluate $\delta T^{\mu\nu}_G$.

Since we take the adiabatic expansion in terms of not $k/m$ but $H/m$, the spatial derivatives do or do not change the order $m^{-1}$ depending on the scales of the inhomogeneities. For the large scales such that
\begin{align}
\frac{k^2}{a^2} \ll mH\,, \label{large_scale}
\end{align}
the spatial derivatives acting on the variables do not change the order of $m^{-1}$, i.e.,
\begin{align}
k^2 =\mathcal{O}(m^0)\,,
\end{align}
in which the consistency of the Einstein equation leads to
\begin{align}
\langle \delta T^{\mu\nu}_G \rangle_T=\mathcal{O}(m^0)\,. \label{order_TG_large}
\end{align}
On the other hand, for the small scales 
\begin{align}
\frac{k^2}{a^2} \gtrsim mH \,, \label{small_scale}
\end{align}
the order of $k$ is
\begin{align}
k^2=\mathcal{O}(m)\,.
\end{align}
Then, the graviton energy-momentum tensor is evaluated as
\begin{align}
\langle \delta T^{\mu\nu}_G \rangle_T=\mathcal{O}(m^1)\,. \label{order_TG_small}
\end{align}
As we will show, this classification of the scales corresponds to the scales beyond or below the Jeans scale.

The graviton mass should be $m \gtrsim 10^{-4}$~eV since we have not detected any deviations from the Newtonian gravitational law in the laboratory scales.\footnote{The constraints are obtained from the linearized theory. The precise constraints on the nonlinear bigravity theory are subject to discussion.} In that case the scale $a/k \sim (mH)^{-1/2}$ is quite small compared with the structures of the Universe. Therefore, the case \eqref{small_scale} is irrelevant to the cosmic structure formation. Nevertheless, we shall discuss both scales \eqref{large_scale} and \eqref{small_scale}, for completeness. Furthermore, the discussion about the small scale \eqref{small_scale} will make the properties of the massive graviton condensate clear.

We have ignored sub-subleading terms to derive the equations \eqref{wave_eq} and \eqref{TT_constraint}. Supposing $\delta \varphi_{\mu\nu}$ is of order $\mathcal{O}(m^0)$, these equation showing the order of errors explicitly are written as
\begin{align}
\left( \nablaB{}_{\alpha} \nablaB{}^{\alpha} -m^2 \right) \varphi_{\mu\nu}&= 0+\mathcal{O}(m^0) \,,  
\label{wave_eq_order_m}
\\
\nablaB{}_{\mu}\varphi^{\mu\nu}&=0+\mathcal{O}(m^{-1}) \,, 
\label{transverse_order_m}
\\ 
\varphi^{\alpha}{}_{\alpha}&= 0+\mathcal{O}(m^{-2})\,.
\label{traceless_order_m}
\end{align}
The corrections to the equations come from the approximations Eqs. \eqref{commute} and \eqref{free_propagate} as well as the nonlinear quantities of $\varphi_{\mu\nu}$.
It is worth noting that, as long as the expressions are linear in inhomogeneities, the accuracy of the calculations are same in both cases \eqref{large_scale} and \eqref{small_scale}.
As for the graviton energy-momentum tensor, although we have ignored higher order corrections to $\delta T^{\mu\nu}_G$ such as $m^2 \bar{\varphi}^3 \delta \varphi$, they are of order $\mathcal{O}(m^{-1})$. Up to the order $\mathcal{O}(m^{0})$, the higher order terms are negligible.

In the standard cosmological perturbation theory, general perturbations can be decomposed into scalar type, vector type, and tensor type perturbations and they are decoupled at the linear order due to the spatial homogeneity and isotropy of the background. In the present case, the background configuration $\bar{\varphi}_{\mu\nu}$ breaks the spatial rotational symmetry and then the different modes could couple in the scalar-vector-tensor decomposition (see \cite{Pereira:2007yy} for perturbations around the anisotropic universe).
However, there still exists the rotational symmetry in the $y-z$ plane. The perturbations can be decomposed into the even parity perturbations and the odd parity perturbations associated with the rotation in the $y-z$ plane:
\begin{align}
\delta g_{\mu\nu}&=\delta g_{\mu\nu}^{\rm (even)}+\delta g_{\mu\nu}^{\rm (odd)}
\,, \\
\delta \varphi_{\mu\nu}&=\delta \varphi_{\mu\nu}^{\rm (even)}+\delta \varphi_{\mu\nu}^{\rm (odd)}
\,,
\end{align}
where explicit forms of perturbations are shown in Appendix \ref{sec_general_pert}. Furthermore, due to the spatial translation symmetry of the background all variables can be transformed into the momentum space, e.g.,
\begin{align*}
\phi(t,\mathbf{x})\rightarrow \phi(t) e^{i (k_x x + k_y y + k_z z)}
\,,
\end{align*}
and the different momentum modes do not couple. Henceforth we use variables in the momentum space and the notation
\begin{align}
k_{\|}^2&=k_y^2+k_z^2\,, \\
k^2&=k_x^2+k_y^2+k_z^2\,.
\label{def_k}
\end{align}

For the calculations, we decompose the perturbations into the odd parity perturbations and the even parity perturbations. However, to clarify the physical meaning of the results, we shall divide the odd parity perturbations and the even parity perturbations into the scalar type, the vector type, and the tensor type components. We define three dimensional harmonic scalar $Y_S$, vectors $Y_V^i, \mathcal{Y}_V^i$, and tensors $Y_T^{ij},\mathcal{Y}_T^{ij}$ which satisfy
\begin{align}
\partial^2 Y_S&=-k^2 Y_S \,, \nn
\partial^2 Y_V^i&=-k^2 Y_V^i\,, \quad \partial^2\mathcal{Y}_V^i=-k^2 \mathcal{Y}_V^i
\,,
\nn
\partial^2 Y_T^{ij}&=-k^2 Y_T^{ij}\,, \quad \partial^2 \mathcal{Y}_T^{ij}=-k^2 \mathcal{Y}_T^{ij}
\,, \label{SVT_harmonics1}
\end{align}
and
\begin{align}
\partial_i Y^i_V&=\partial_i \mathcal{Y}^i_V=0\,,
\nn
\partial_i Y^{ij}_T&=\partial_i \mathcal{Y}^{ij}_T=0\,, \quad
Y^i_T{}_i=\mathcal{Y}^i_T{}_i=0
\,, \label{SVT_harmonics2}
\end{align}
where $\partial^2=\partial_i \partial^i$ and $i,j=(x,y,z)$. The indices $i,j$ are raised and lowered by $\delta^{ij}$ and $\delta_{ij}$. The quantities $Y_S,Y_V^i,Y_T^{ij}$ are even parity quantities associated with the two dimensional rotation in the $y-z$ plane while $\mathcal{Y}_V^i,\mathcal{Y}_T^{ij}$ are odd parity quantities. The suffixes $S,V$, and $T$ are attached to classify the quantities into the scalar, the vector, and the tensor types associated with the three dimensional rotation.
We further introduce the quantities as
\begin{align}
Y_S^i&=-\frac{1}{k}\partial^iY_S
\,, \\
Y_S^{ij}&=k^{-2}\left( \partial^i \partial^j-\frac{1}{3}\partial^k \partial_k \delta^{ij} \right)Y_S
\,, \\
Y_V^{ij}&=-\frac{1}{k}\partial^{(i}Y_V^{j)}\,, 
\\
\mathcal{Y}_V^{ij}&=-\frac{1}{k}\partial^{(i}\mathcal{Y}_V^{j)}
\,,
\end{align}
then we obtain nine harmonics which are summarized in Table \ref{harmonics}

\begin{table}[h]
\caption{The classifications of the even parity perturbations and the odd parity perturbations.}
\label{harmonics}
\begin{tabular}{ccccc}
\hline
            & scalar & vector & tensor \\
\hline\hline
even parity \,&\, $Y_S,Y_S^i, Y_S^{ij}$ \,&\, $Y_V^i, Y_V^{ij}$ \,&\, $Y_T^{ij}$ \\
odd  parity & none   & $\mathcal{Y}_V^i, \mathcal{Y}_V^{ij}$ & $\mathcal{Y}_T^{ij}$ \\
\hline
\end{tabular}
\end{table}

Using the gauge condition (see Appendix \ref{sec_general_pert} for details), the perturbations for the low-frequency mode $\delta g_{\mu\nu}$ can be given by
\begin{align}
\delta g^{\rm even}_{\mu\nu}&=
\begin{pmatrix}
-2\Phi Y_S & -a B_V Y_V{}_j \\
* & 2a^2 (\Psi Y_S \delta_{ij}+H_T Y_T{}_{ij})
\end{pmatrix}
\,, \label{delg_even}\\
\delta g^{\rm odd}_{\mu\nu}&=
\begin{pmatrix}
0 & -a \mathcal{B}_V \mathcal{Y}_V{}_j \\
* & 2a^2 \mathcal{H}_T \mathcal{Y}_T{}_{ij}
\end{pmatrix}
\,, \label{delg_odd}
\end{align}
where $\Phi,\Psi,B_V,H_T,\mathcal{B}_V,\mathcal{H}_T$ are functions of $t$.

We will calculate the perturbed graviton energy-momentum tensor in the following subsections. In the momentum space, the perturbed graviton energy-momentum tensor is given by the form
\begin{widetext}
\begin{align}
\langle \delta T_G^{\rm (even)}{}^{\mu\nu} \rangle_T&=
\begin{pmatrix}
(\delta \rho_G -2\Phi \bar{\rho}_G) Y_S & a^{-1}\bar{\rho}_G v_{\rm (even)}^j \\
* & a^{-2}\left[\delta p_G \, Y_S  \delta^{ij} +\Pi^{ij}_{\rm (even)}\right]
\end{pmatrix}
\,, \label{TG_even}
\\
\langle \delta T_G^{\rm (odd)}{}^{\mu\nu} \rangle_T&=
\begin{pmatrix}
0 & a^{-1}\bar{\rho}_G v_{\rm (odd)}\mathcal{Y}_V^j \\
* & a^{-2}\Pi^{ij}_{\rm (odd)}
\end{pmatrix}
\,, \label{TG_odd}
\end{align}
\end{widetext}
from which we can read the energy, the velocity, the isotropic pressure, and the anisotropic stress regarding the massive graviton condensate as a fluid. The velocity for the even parity perturbation is decomposed into
\begin{align}
v^i_{\rm (even)}=v_S Y_S^i+v_V Y_V^i\,,
\end{align}
whereas the anisotropic stresses are decomposed into
\begin{align}
\Pi_{\rm (even)}^{ij}&=\pi_S Y_S^{ij}+\pi_V Y^{ij}_V+\pi_T Y^{ij}_T\,,
\\
\Pi_{\rm (odd)}^{ij}&=\pi^{\rm (odd)}_V \mathcal{Y}_V^{ij}+\pi^{\rm (odd)}_T \mathcal{Y}^{ij}_T\,.
\end{align}

Since calculations for general perturbations are complicated, we just show particular solutions. The generic solutions are given in Appendix \ref{sec_general_pert}.

First, the odd parity perturbations are not important for the structure formation since these modes contain only the vector type and the tensor type perturbations. We discuss only the even parity perturbations in this section. Furthermore, we find the vector type components are always decoupled up to the subleading order whereas the scalar type and the tensor type components are coupled in the small scales. Since the vector type perturbations represent the rotational modes and decay in time, the vector modes are not important. Hence, it is sufficient for the structure formation to consider the even parity perturbations without the contributions from the vector type components. 

The irrotational solution for the even parity perturbations can be found under the ansatz
\begin{widetext}
\begin{align}
\delta g_{\mu\nu}(t,\mathbf{k})&=
\begin{pmatrix}
-2\Phi Y_S & 0 \\
* & 2a^2 (\Psi Y_S \delta_{ij}+H_T Y_T{}_{ij})
\end{pmatrix}
\,, \label{even_gmunu} \\
\delta \varphi_{\mu\nu}(t,\mathbf{k}) &=
\begin{pmatrix}
-2 \phi Y_S  & -a B Y_x & -a C Y_a \\
* & 2a^2(\psi+ 2\delta \varphi) Y_S & 0 \\
* & * & 2a^2 (\psi-\delta \varphi) \delta_{ab} Y_S
\end{pmatrix}
+\mathcal{O}(m^{-2})
\,, \label{large_varphi}
\end{align}
\end{widetext}
where $a,b=(y,z)$ whose indices are raised and lowered by $\delta_{ab}$ and $Y_x$ and $Y_a$ are defined by \eqref{even_harmonics}.
Since the non-diagonal components of $\delta \varphi_{ij}$ only have decaying modes, we just set $\delta \varphi_{ij}=0$ for $i\neq j$.
Note that we do not use the scalar-vector-tensor type harmonics to represent the components of the massive graviton in order that the expression is written in a clear form. Under the adiabatic expansion, the variables for the massive graviton can be given by
\begin{align}
\phi&=\phi_1 \cos[mt]+\phi_2 \sin[mt] \,, \nn
B&=B_1 \cos[mt]+B_2 \sin [mt] \,, \nn
C&=C_1 \cos[mt]+C_2 \sin [mt] \,, \nn
\psi &=\psi_1 \cos[mt]+\psi_2 \sin [mt] \,, \nn
\delta \varphi &=\delta \varphi_1 \cos[mt]+\delta \varphi_2 \sin [mt] \,, 
\label{slowly_functions}
\end{align}
with slowly varying functions of time $\{ \phi_{1,2},B_{1,2},C_{1,2},\psi_{1,2},\delta \varphi_{1,2}\}$. We shall show the particular solutions for the large scales \eqref{large_scale} and the small scales \eqref{small_scale} in order.

\subsection{Large scale inhomogeneity}
In the large scales \eqref{large_scale}, the consistency of the equations leads to that the amplitudes of the variables have to be
\begin{align}
\Phi,\Psi,H_T,\delta \varphi_{2}&=\mathcal{O}(m^{0}) \,,\nn
B_1,C_1,\psi_1,\delta \varphi_1 &=\mathcal{O}(m^{-1}) \,, \nn
\phi_{1,2},B_2,C_2,\psi_2 &=\mathcal{O}(m^{-2})
\,.
\end{align}
The order $\mathcal{O}(m^{-2})$ quantities give just sub-subleading contributions thus we can ignore them.

Eqs.~\eqref{transverse_order_m} and \eqref{traceless_order_m} yield the constraint equations which determine $B_1,C_1$ and $\psi_1$ as
\begin{align}
B_1=-\frac{4k}{ma}\delta \varphi_2 \,, \, C_1=\frac{2k}{ma}\delta \varphi_2 \,, \,
\psi_1=\frac{2k_{\|}^4}{k^4}H_T \bar{\varphi}_1
\,.
\end{align}
Eq.~\eqref{wave_eq_order_m} gives
\begin{align}
\delta \dot{\varphi}_2+\frac{3}{2}H \delta \varphi_2+ m \Phi \bar{\varphi}_1=0
\,.\label{varphi2}
\end{align}
We do not find other equations within our accuracy.

After using the above equation and taking the time average, we obtain
\begin{align}
\delta \rho_G&=6m^2 \left[\bar{\varphi}_1 \delta \varphi_1+\bar{\varphi}_2 \delta \varphi_2+\bar{\varphi}_1^2  \left(2\Psi+\frac{k_{\|}^4}{k^4}H_T \right) \right]
, \label{rhoG_large} \\ 
v_S&=-\frac{k}{am \bar{\varphi}_1}\delta \varphi_2 \label{vS_large}
\,, \\
v_V&, \delta p_G, \pi_S, \pi_V,\pi_T=0
\,,
\end{align}
from the perturbed graviton energy-momentum tensor. Therefore, the graviton energy-momentum tensor is given by a form of pressureless perfect fluid without the vector type components (i.e., an irrotational fluid). Although the irrotational property is obtained because of our specific ansatz \eqref{large_varphi}, the pressureless property is hold even if we consider the general solution.

Note that the evolution of $\delta \varphi_1$ (and also $\bar{\varphi}_2$) has not been determined within our accuracy. We can, however, determine the dynamics of $\delta \rho_G$ by choosing the combinations $\delta \rho_G$ as an independent variable instead of $\delta \varphi_1$. The dynamics of $\delta \rho_G$ is determined by the conservation law of the averaged graviton energy-momentum tensor\footnote{We notice again that although the conservation law of the non-averaged graviton energy-momentum tensor $\nablaB_{\mu} T^{\mu\nu}_G=0$ is equivalent to the equation of motion of $\varphi_{\mu\nu}$, that of the averaged graviton energy-momentum tensor $\nablaB_{\mu} \langle T^{\mu\nu}_G \rangle_T=0$ is not. Hence, we can obtain the equation for $\delta \rho_G$ even if the equations for $\delta \varphi_1$ cannot be obtained from the equation of motion within our accuracy.}
which reads
\begin{align}
\delta \dot{\rho}_G +3H\delta \rho_G+\left(\frac{k}{a}v_S+3\dot{\Psi} \right) \bar{\rho}_G &=0
\,, \label{eq_delta_rho} \\
\dot{v}_S+H v_S-\frac{k}{a}\Phi &=0
\,, \label{eq_vS}
\end{align}
where we notice that \eqref{eq_vS} is exactly same as Eq.~\eqref{varphi2}. Eq.~\eqref{eq_delta_rho} determines the dynamics of $\delta \rho_G$.

Although the definition of $\delta \rho_G$ contains the tensor mode $H_T$, Eq.~\eqref{eq_delta_rho} indicates that the dynamics of $\delta \rho_G$ is independent from the tensor mode $H_T$. Therefore, if we focus on only the macroscopic behaviour of the massive graviton condensate (i.e., we focus on only the dynamics of $\langle T^{\mu\nu}_G \rangle_T$), the scalar modes and the tensor mode are decoupled. Needless to say, at the microscopic level, the scalar modes and the tensor modes should be coupled. For example, the dynamics of $\delta \varphi_1$ would depend on the dynamics of the tensor mode $H_T$ as well as the scalar modes $\Phi,\Psi$. Furthermore, the couplings between the scalar-vector-tensor modes would appear when we consider more higher order corrections. The present calculations are justified up to the sub-subleading order.

As a result, the massive graviton condensate behaves as a dust fluid in the large scales \eqref{large_scale} even if a small inhomogeneity is introduced. The massive graviton condensate can cluster and then explain the cosmic structure formation.

\subsection{Small scale inhomogeneity}
Next, we discuss the solution in the small scales \eqref{small_scale} in which the amplitudes are given by
\begin{align}
\Phi,\Psi,H_T,\delta \varphi_{1,2}&=\mathcal{O}(m^0) \,, \nn
B_{1,2},C_{1,2}&=\mathcal{O}(m^{-1/2}) \,, \nn
\phi_{1,2},\psi_{1,2}&=\mathcal{O}(m^{-1})
\,.
\end{align}
The constraint equations \eqref{transverse_order_m} and \eqref{traceless_order_m} yield
\begin{align}
\phi_1+3\psi_1-6\frac{k_{\|}^4}{k^4}H_T \bar{\varphi}_1&=0
\,, \nn
\phi_2+3\psi_2&=0
\,, \nn
\pm \frac{2mk}{a}\phi_{1,2}+\frac{k_x^2}{a^2}B_{2,1}+\frac{k_{\|}^2}{a^2}C_{2,1}&=0
\,,
\end{align}
and
\begin{align}
B_{1,2}=\mp \frac{4k}{ma}\delta \varphi_{2,1} \,, \quad C_{1,2}=\pm \frac{2k}{ma}\delta \varphi_{2,1} 
\,,
\end{align}
whereas \eqref{wave_eq_order_m} gives
\begin{align}
\delta \dot{\varphi}_1+\frac{3}{2}H \delta \varphi_1 -\frac{k^2}{2ma^2}\delta \varphi_2 &=0
\,, \label{eq_varphi1} \\ 
\delta \dot{\varphi}_2+\frac{3}{2}H \delta \varphi_2 +\frac{k^2}{2ma^2}\delta \varphi_1+m \Phi \bar{\varphi}_1&=0
\,. \label{eq_varphi2}
\end{align}
Note that there are ten independent equations for ten independent variables $\{ \delta \varphi_{1,2}, B_{1,2},C_{1,2},\phi_{1,2},\psi_{1,2}\}$. Hence, the dynamics of the massive graviton are completely determined within our accuracy, differently from the previous case.

We notice that the equations \eqref{eq_varphi1} and \eqref{eq_varphi2} yield the Schr\"odinger equation in the cosmological background. The ``wavefunction'' $u(t,\mathbf{x})$ is defined by the relation
\begin{align}
\bar{\varphi}(t)+\delta \varphi(t,\mathbf{x})  = \frac{1}{2} \left[ u(t,\mathbf{x}) e^{-imt}+u^*(t,\mathbf{x}) e^{imt} \right]
\,,
\end{align}
where $\delta \varphi(t,\mathbf{x})$ is the variable in the real space which is used only here. Then, we obtain
\begin{align}
i \left( \frac{\partial u}{\partial t}+\frac{3}{2}Hu \right)= \left( -\frac{\partial^2 }{2m^2 a^2}+m\Phi  \right) u\,,
\label{Schrodinger}
\end{align}
where $m\Phi u \simeq m \Phi \bar{\varphi}_1$ since we have considered the linearized theory.
The wavefunction $u$ is dominated by the coherent mode $u_0=\bar{\varphi}_1$ which suggests that the almost homogeneous configuration of $\varphi_{\mu\nu}$ represents the condensate of the massive graviton. A large fraction of massive gravitons occupies the state $u_0$ except for the tiny perturbations $\delta \varphi_{\mu\nu}$. Note that $u_0$ is also a solution to the Schr\"odinger equation since the gravitational potential $\Phi$ is zero for the homogeneous configuration.

We obtain
\begin{align}
\delta \rho_G &=6m^2 \bar{\varphi}_1 \delta \varphi_1 +\mathcal{O}(m^0)
\,, \label{rhoG_small} \\
v_S&=-\frac{k}{am\bar{\varphi}_1}\delta \varphi_2
\,, \label{vS_small} \\
\delta p_G&=-\frac{\bar{\varphi}_1}{6a^2}
 \left( 11k_x^2+5 k_{\|}^2 \right)\delta \varphi_1 
\,, \\
\pi_S&=-\frac{\bar{\varphi}_1}{2a^2} \left(10k_x^2+7k_{\|}^2 \right) \delta \varphi_1
\,, \\
\pi_T&=-\frac{3\bar{\varphi}_1}{2a^2}
  k^2 \delta \varphi_1 \,,
\end{align}
and
\begin{align}
v_V=\pi_V=0
\,.
\end{align}
Therefore, the massive graviton condensate is no longer recognized as a pressureless fluid. The effect of the pressure is relevant within the Jeans length. The Jeans momentum $k_J$ is estimated by the relation
\begin{align}
\left| \frac{k^2}{a^2} \frac{ \delta p_G}{\delta \rho_G} \right|_{k=k_J} \sim \frac{\bar{\rho}_G}{M_{\rm pl}^2}\,,
\end{align}
which yields
\begin{align}
\frac{k_J^2}{a^2} \sim m H\,.
\end{align}
where we have assumed $M_{\rm pl}^2H^2\sim \bar{\rho}_G$.
Hence, \eqref{small_scale} indeed correspond to the scales below the Jeans scale. 

If the massive graviton condensate is the dominant component, the Einstein equation yield
\begin{align}
\frac{k^2}{a^2}\Psi =\frac{1}{2M_{\rm pl}^2} \delta \rho_G
\,, \quad
\frac{k^2}{a^2}(\Phi+\Psi)=-\frac{\pi_S}{M_{\rm pl}^2}
\,.
\end{align}
Since $\pi_S$ is of order $\mathcal{O}(m^0)$, we obtain $\Psi=-\Phi+\mathcal{O}(m^{-1})$. Eq.~\eqref{eq_varphi2} then becomes
\begin{align}
\delta \dot{\varphi}_2+\frac{3}{2}H \delta \varphi_2+\left( \frac{k^2}{2ma^2}-\frac{ma^2}{k^2} \frac{\bar{\rho}_G}{M_{\rm pl}^2} \right) \delta \varphi_1=0\,.
\end{align}
Combining with Eq.~\eqref{eq_varphi1}, we find
\begin{align}
\ddot{\delta}+2H \dot{\delta}+\left( \frac{k^4}{4m^2 a^4}-\frac{\bar{\rho}_G}{2M_{\rm pl}^2} \right) \delta =0
\,,
\end{align}
where $\delta :=\delta \rho_G/\bar{\rho}_G=2\delta \varphi_1/\bar{\varphi}_1$ is the relative perturbation of the energy density. Clearly, in the scales beyond the Jeans scale, this equation admits a growing mode solution $\delta \propto a$ due to the Jeans instability in the dust dominant universe. On the other hand, we find
\begin{align}
\delta \propto \exp \left[\pm i\frac{3k^2}{2ma^2} t \right]\,,
\end{align}
for $k^2 \gg k_J^2$ with $a\propto t^{2/3}$. The massive graviton condensate shows the acoustic oscillation.

The tensor mode is not decoupled from the scalar mode. The gravitational wave $H_T$ is sourced by the anisotropic stress $\pi_T$ which is related with the energy density of the massive graviton condensate. The acoustic oscillation of the massive graviton emits the gravitational waves.


\section{Production of massive graviton condensate}
\label{production}
As shown in the previous section, the massive graviton condensate is indeed a candidate of dark matter. We thus consider a production mechanism of the massive graviton condensate and discuss whether the massive graviton condensate can be the dominant component of dark matter.

To generate the massive graviton condensate we need an anisotropic source which is coherent on the cosmological scale\footnote{Although we have assumed that the massive graviton is coherent on the cosmological scale for simplicity, we can discuss the case when the coherent scale is smaller than the horizon scale but larger than, at least, the de Broglie wavelength of the massive graviton. The source is not necessary to be anisotropic over the cosmological scale.}. A candidate is the cosmological scale magnetic field. The blazar observations suggests the lower bound of the strength of the extragalactic magnetic field $B_0$ is about $10^{-17}$ G. The upper bound is obtained from the CMB observations which is about $10^{-9}$ G \cite{Ade:2015cva}. Since the production and the evolution of the cosmological scale magnetic field are subject to discussion (see \cite{Kandus:2010nw,Durrer:2013pga,Subramanian:2015lua} for reviews), we consider the simplest scenario.

We assume that the coherent magnetic field is generated in the early universe, e.g., in the inflationary regime of the universe. Here, we do not discuss the initial spectrum of the dark matter perturbations which should depend on the details of the coherent magnetic field. We just estimate the produced amount of the coherent massive gravitons. If the magnetic field adiabatically evolves, the energy density decreases as $a^{-4}$. On the other hand, the energy density of the massive graviton decreases as $a^{-3}$. Therefore, the production of the massive graviton condensate by the magnetic field can be ignored in the late stage of the universe $(m \gg H)$. 

\begin{figure}[htbp]
\centering
\includegraphics[width=6.5cm,angle=0,clip]{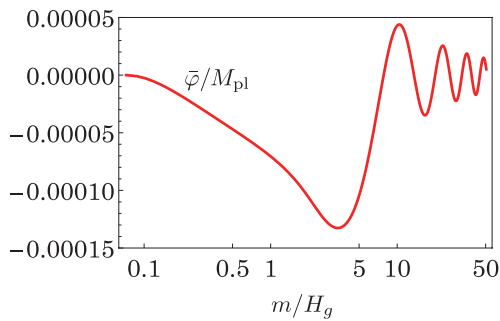}
\caption{A production of $\bar{\varphi}$ by the coherent magnetic field with $\Omega_B/\Omega_r=10^{-4}$. $H_g$ is the Hubble expansion rate of the $g$-spacetime which becomes $H_g \simeq H$ in $H_g \ll m$. We set $b_1=-1$ and initially set both the $g$-spacetime and the $f$-spacetime are isotropic, i.e., $\bar{\varphi}=0$ and $\dot{\bar{\varphi}}=0$.
}
\label{production_fig}
\end{figure}

The separation \eqref{low+high} is not justified in $H \gtrsim m$. To discuss the early universe we have to directly analyze the Bianchi universe which is summarized in Appendix \ref{sec_Bianchi}. By solving Eqs.~\eqref{BianchiI_Fri}-\eqref{eq_sigmaf}, a typical behavior of $\bar{\varphi}$ produced by the coherent magnetic field is shown in Fig.~\ref{production_fig}.

The amplitude of $\varphi_{\mu\nu}$ does not grow in $H \gg m$ since the Hubble friction is too large. As a result, the dominant production of the massive graviton condensate should occur at $H\sim m$. The produced amplitude of the condensate is estimated as
\begin{align}
|\bar{\varphi}_*| \sim \frac{B_*^2}{M_{\rm pl}m^2} \sim  \frac{B_*^2}{M_{\rm pl} H_*^2}   \sim M_{\rm pl} \frac{\Omega_B}{\Omega_r}
\,, \label{produced_phi}
\end{align}
where the asterisk represents the quantities at the production and $B$ is the strength of the magnetic field. $\Omega_B$ and $\Omega_r$ are the present density parameters of the coherent magnetic field and radiation, respectively. To obtain the final expression we use the universe is dominated by radiation at the production, $M_{\rm pl}^2H_*^2 \sim \rho^*_r$, and both $B^2$ and $\rho_r$ decrease as $a^{-4}$. Once the gravitons are produced, $\varphi_{\mu\nu}$ is decoupled from the magnetic field because the contribution from the interaction quickly decreases (see Eq.~\eqref{Bianchi_phi}). In order that $\varphi_{\mu\nu}$ is the dominant component of dark matter, the present amplitude has to be $\bar{\varphi}_0 \sim M_{\rm pl} H_0/m$ where the present and the produced amplitudes are related by $a_0^{3/2}\bar{\varphi}_0 = a_*^{3/2}\bar{\varphi}_*$. Then, we get the condition for the graviton mass being the dominant component of dark matter as
\begin{align}
m &\sim H_0 z_{\rm eq}^{-3/2} \left(\frac{\Omega_r}{\Omega_B} \right)^4
\nn 
&\sim 10^{-4} \left( \frac{10^{-10} {\rm G}}{B_0} \right)^8 {\rm eV}
\,.
\end{align}
where $z_{\rm eq}$ is the redshift of the equality time.
The graviton mass should be $10^{-4} {\rm eV} \lesssim m \lesssim 10^7 {\rm eV}$ where the lower bound is obtained from laboratory-scale experiments of gravity while the upper bound is given by the lifetime of the massive graviton. Hence, a consistent scenario is constructed when the present magnetic field is
\begin{align}
10^{-12} G \lesssim B_0 \lesssim 10^{-10} G
\,,\label{magnetic_field}
\end{align}
which is indeed a viable region of the coherent magnetic field.

\section{Summary and discussions}
\label{summary}
We provide a scenario in which a tiny deformation of the spacetime is dark matter in the ghost-free bigravity theory. This deformation is interpreted as the ``condensate'' of the massive gravitons. Differently from the case of the massless graviton, the zero momentum state of the massive graviton is well-defined when $m^2 \gg H^2$. We find that the zero momentum massive graviton with small fluctuations is a viable candidate of dark matter.

We have also studied a production mechanism of the coherent massive gravitons with the mass range $10^{-4}~{\rm eV} \lesssim m \lesssim 10^7$~eV and shown the coherent magnetic field with $10^{-12}~{\rm G} \lesssim B_0 \lesssim 10^{-10}~{\rm G}$ yields a sufficient amount of massive gravitons for dark matter. When the present value of the coherent magnetic field is determined by a future observation, we can fix a suitable value of the graviton mass to be dark matter.

Although we discussed the magnetic field as a source of the massive graviton condensate, another source to produce the condensate could exist. In general, if there exits a coherent anisotropic stress $\pi_{\rm coh}$ whose coherent scale is $L\gtrsim m^{-1}$ and the density is $\pi_{\rm coh}/\rho_r \sim 10^{-10}$ in the age $H\sim m$, the massive graviton condensate is produced and becomes dark matter.  Since the gravitons universally couple to matter fields, the source is not necessary to be a standard model particle. Any matter field can be a source of the gravitons.

If the anisotropic stress is a random field instead of the coherent field, the stochastic massive gravitons are produced which have been discussed in \cite{Aoki:2016zgp}. Even for the stochastic case, we obtain a viable scenario of the massive graviton dark matter. Hence, if the anisotropic stress existed in the early universe, it inevitably yields the stochastic or the coherent massive gravitons and then the massive gravitons can be dark matter.

One may interpret that the $Z_2$ symmetry is a fundamental symmetry of the massive graviton. Although we have assumed the $Z_2$ symmetry only for the self-interactions of gravitons, the symmetry can be introduced into the matter-graviton interactions as well. The $Z_2$ symmetry holds if the action is invariant under the replacement $g \leftrightarrow f$ in bigravity. At low energy scales, matter fields can couple to both metrics $g_{\mu\nu}$ and $f_{\mu\nu}$ via an effective composite metric~\cite{Yamashita:2014fga,deRham:2014fha,deRham:2014naa,Heisenberg:2014rka,Hinterbichler:2015yaa,Heisenberg:2015iqa,deRham:2015cha,Huang:2015yga,DeFelice:2015yha}. Hence, when all matter fields couple to the composite metric $g^{\rm eff}_{\mu\nu}$ defined by
\begin{align}
g^{\rm eff}_{\mu\nu}=\frac{1}{4}\left[g_{\mu\nu}+2 \left(\sqrt{gf}\right)_{\mu\nu} +f_{\mu\nu} \right]\,,
\end{align}
with
\begin{align}
\left(\sqrt{gf} \right)_{\mu\nu}=g_{\mu\alpha} \left(\sqrt{g^{-1}f}\right)^{\alpha}{}_{\nu}\,,
\end{align}
the matter-massive graviton interactions respect the $Z_2$ symmetry. Indeed, expanding the metrics under \eqref{low+high} we obtain
\begin{align}
g^{\rm eff}_{\mu\nu}=\gB{}_{\mu\nu}+\frac{h_{\mu\nu}}{M_{\rm pl}}-\frac{1}{4M_{\rm pl}^2}\varphi_{\mu\alpha} \varphi^{\alpha}{}_{\nu} +\cdots
\,,
\end{align}
thus, the matter action is manifestly invariant under $\varphi_{\mu\nu} \rightarrow -\varphi_{\mu\nu}$. In this case, the Yukawa interaction of the massive graviton does not appear and then the present graviton mass constraints cannot be applied. The details about the $Z_2$ symmetric bigravity theory are under investigation.

We have introduced the $Z_2$ symmetry to the self-interactions of the massive graviton in order to simplify the calculations. However, the massive graviton condensate can be dark matter even without the $Z_2$ symmetry because the nonlinear terms are always sub-leading contributions and then may not affect the dynamics at leading order. The leading order expression of $T^{\mu\nu}_G$ would be unchanged. In order that the massive graviton condensate is dark matter, the $Z_2$ symmetry of the self-interactions should not be required.

Our scenario can be directly confirmed when we observe the coherent anisotropic oscillation of the Universe. The frequency of the oscillation is unfortunately too high to detect the oscillation as a ``gravitational wave'' by the present and future gravitational wave detectors. However, if the graviton mass can be sufficently light due to, for example, the $Z_2$ symmetry, the coherent oscillation will be detectable.

Finally, we comment on an interesting remaining question: Is the almost homogeneous configuration of $\varphi_{\mu\nu}$ the Bose-Einstein condensate of the massive graviton? As is well-known, a coherent massive scalar field, for example axion, is a viable dark matter candidate~\cite{Turner:1983he,Press:1989id,Sin:1992bg,Ji:1994xh,Lee:1995af,Hu:2000ke,Goodman:2000tg,Peebles:2000yy,Amendola:2005ad,Boehmer:2007um,Lee:2008jp,Sikivie:2009qn,Chavanis:2011uv,Schive:2014dra,Hui:2016ltb}. This coherent scalar field is interpreted as the Bose-Einstein condensate. Our result would be a generalization of the Bose-Einstein condensate of the massive scalar field to that of the massive tensor field. However, the present argument is completely classical and we have not discussed any quantum aspect of the massive graviton. Therefore, it would be interesting to study a connection to the quantum theory of the gravitation, but this is beyond the scope of the present paper.

\section*{Acknowledgments}
K.A. would like to thank Shinji Mukohyama and Ryo Namba for useful discussions and comments.  This work was supported in part by Grants-in-Aid from the Scientific Research Fund of the Japan Society for the Promotion of Science  (No. JP15J05540, No. JP16K05362, and No. JP17H06359). 


\appendix
\section{Graviton energy-momentum tensor: curved background}
\label{sec_Tmunu}

In this section, we summarize the definitions of the graviton energy-momentum tensors for generic cases. For completeness, we introduce both the $g$-matter fields $\psi_g$ and the $f$-matter fields $\psi_f$ and do not assume \eqref{symmetric_condition}. The matter action is given by
\begin{align}
S^{\rm [m]}=S^{\rm [m]}_g(g,\psi_g)+\mathcal{S}^{\rm [m]}_f(f,\psi_f)
\,\,
\end{align}
and we denote the energy-momentum tensors of the $g$-matter field and the $f$-matter field as $T^{\mu\nu}$ and $\mathcal{T}^{\mu\nu}$, respectively.

To define the energy-momentum tensor of gravitons, we shall decompose the metric into the ``background'' and the ``perturbations''. However, in general, the deomposition into the perturbations and the background may not be well-defined because the perturbations and the background interact with each others and then the equations may not be separable when the backreaction from the perturbations to the background is included. To decompose the Einstein equations, we assume the perturbations contain only high-frequency modes whereas the background consists of only the low-frequency modes. In this case, we obtain two independent eqautions for the low-frequency background and the high-frequency perturbations via a low-frequency projection $\langle \cdots \rangle_{\rm low}$ and a high-frequency projection $\langle \cdots \rangle_{\rm high}$, respectively. Therefore, we assume the metrics are expressed by the low-frequency backgrounds with the high-frequency perturbations:
\begin{align}
g_{\mu\nu}=g_{\mu\nu}^{\rm (low)}+\delta g_{\mu\nu}^{\rm (high)} \,, \\
f_{\mu\nu}=f_{\mu\nu}^{\rm (low)}+\delta f_{\mu\nu}^{\rm (high)} \,,
\end{align}
with $|\delta g^{\rm (high)}_{\mu\nu}| \ll |g^{\rm (low)}_{\mu\nu}|$ and $|\delta f^{\rm (high)}_{\mu\nu}| \ll |f^{\rm (low)}_{\mu\nu}|$. The high-frequency mode and the low-frequency mode are defined by
\begin{align}
g^{\rm (low)}_{\mu\nu}=\langle g_{\mu\nu} \rangle_{\rm low}\,,\quad
\delta g^{\rm (high)}_{\mu\nu}=\langle g_{\mu\nu} \rangle_{\rm high}\,,
\end{align}
which is same for $f_{\mu\nu}$.
It is worth noting that we only assume the perturbations are high-frequency modes but do not assume the perturbations are high-momentum modes. In GR, the situations with the high-frequency waves and the situations with the high-momentum waves are equivalent since the graviton is massless. However, these situations are not equivalent in bigravity due to the existence of the massive graviton.

As already mentioned, the definitions of the massless mode and the massive mode are ambiguous in general. They can be defined when the curvature scale of the metrics are smaller than the graviton mass squared
\begin{align}
|\partial_{\alpha} \partial_{\beta} g_{\mu\nu}| \ll m^2 \,, \quad |\partial_{\alpha} \partial_{\beta} f_{\mu\nu} | \ll m^2 \,.
\end{align}
In this case, although the low-frequency modes of the massive gravitons can be excited by some source including matter as well as the backreactions from high-frequency modes, the amplitudes of the low-frequency massive modes must be tiny (Indeed, in Appendix \ref{sec_Bianchi}, we will see the amplitude of the low-frequency massive mode is suppressed by $m^{-2}$ for the homogeneous matter distributions.). Hence, we may assume the background is approximated by the homothetic solution. Then, the spacetime are expressed as
spacetimes are expressed as
\begin{align}
g_{\mu\nu}&=\gB{}_{\mu\nu}+M_{\mu\nu} +\frac{h_{\mu\nu}}{M_{\rm pl}}+\frac{\varphi_{\mu\nu}}{M_G} \,, 
\label{g_spacetime} \\
f_{\mu\nu}&=\xi_0^2 \left( \gB{}_{\mu\nu}-\alpha^{-1} M_{\mu\nu} +\frac{h_{\mu\nu}}{M_{\rm pl}}-\frac{\varphi_{\mu\nu}}{\alpha M_G}\right)
\,, \label{f_spacetime}
\end{align}
where $\gB{}_{\mu\nu}$ and $M_{\mu\nu}$ are the low-frequency massless mode and the low-frequency massive mode while $h_{\mu\nu}$ and $\varphi_{\mu\nu}$ are the high-frequency massless mode and the low-frequency massive mode, respectively (see Table \ref{table_metric}). The assumption that the homothetic background is good approximation means $|M_{\mu\nu}| \ll |\gB{}_{\mu\nu}|$. In this case, $M_{\mu\nu},h_{\mu\nu}$ and $\varphi_{\mu\nu}$ can be treated as the tensors with respect to $\gB{}_{\mu\nu}$.

\begin{widetext}
The Ricci tensor for the $g$-spacetime is expanded as
\begin{align}
R_{\mu\nu}&=\RB{}_{\mu\nu}+\delta \Rf{}_{\mu\nu}[M]+\delta \Rf{}_{\mu\nu}[\delta g^{\rm (high)}] 
+\delta \Rff{}_{\mu\nu}[\Delta g]+\cdots\,,
\end{align}
where $\Delta g_{\mu\nu}=M_{\mu\nu}+\delta g^{\rm (high)}_{\mu\nu}$. The high-frequency modes of $g_{\mu\nu}$ and $f_{\mu\nu}$ are written in terms of the mass eigenstates as 
\begin{align}
\delta g_{\mu\nu}^{\rm (high)}&=\frac{h_{\mu\nu}}{M_{\rm pl}}+ \frac{\varphi_{\mu\nu}}{M_G} \,, \\
\delta f_{\mu\nu}^{\rm (high)}&= \xi_0^2 \left( \frac{h_{\mu\nu}}{M_{\rm pl}}- \frac{\varphi_{\mu\nu}}{\alpha M_G} \right)
\,.
\end{align}
The functionals $\delta \Rf{}_{\mu\nu}[\chi]$ and $\delta \Rff{}_{\mu\nu}[\chi]$ are defined by
\begin{align}
\delta \Rf{}_{\mu\nu}[\chi]&=\frac{1}{2} \left( -\nablaB_{\alpha} \nablaB{}^{\alpha} \chi_{\mu\nu} - \nablaB{}_{\mu} \nablaB{}_{\nu} \chi^{\alpha}{}_{\alpha} +2 \nablaB{}^{\alpha} \nablaB{}_{(\mu} \chi_{\nu)\alpha} \right)\,,
\\
\delta \Rff{}_{\mu\nu}[\chi]&=\frac{1}{2} \gB{}^{\rho\sigma} \gB{}^{\alpha\beta}
\Biggl[ \frac{1}{2} \nablaB{}_{\mu} \chi_{\alpha \rho} \nablaB{}_{\nu} \chi_{\sigma \beta}
+2\nablaB{}_{\rho} \chi_{\nu \alpha} \nablaB{}_{[\sigma} \chi_{\beta ] \mu}
\nn
&\qquad \qquad \qquad
+\chi_{\rho \alpha} \left( \nablaB{}_{\nu} \nablaB{}_{\mu} \chi_{\sigma \beta} +\nablaB{}_{\beta} \nablaB{}_{\sigma} \chi_{\mu\nu} -2\nablaB{}_{\beta} \nablaB{}_{(\mu} \chi_{\nu) \sigma} \right)
\nn
&\qquad \qquad \qquad
+\left( \frac{1}{2} \nablaB{}_{\alpha} \chi_{\rho \sigma} -\nablaB{}_{\rho} \chi_{\alpha\sigma} \right)
\left( 2\nablaB{}_{(\mu}\chi_{\nu) \beta} -\nablaB{}_{\beta} \chi_{\mu\nu} \right) \Biggl]\,,
\end{align}
for a tensor $\chi_{\mu\nu}$. The linear quantities in $M_{\mu\nu}$ and $\delta g_{\mu\nu}^{\rm (high)}$ are decomposed whereas the quadratic quantity $\delta \Rff{}_{\mu\nu}[\Delta g]$ have the cross terms between $M_{\mu\nu}$ and $\delta g_{\mu\nu}^{\rm (high)}$. The quadratic quantity is given by 
\begin{align}
\delta \Rff{}_{\mu\nu}[\Delta g]
&=\delta \Rff{}_{\mu\nu}[\delta g^{\rm (high)}]+\delta \Rff{}_{\mu\nu}[M]
+\delta \Rff{}^{\rm cross}_{\mu\nu}[M\times \delta g^{\rm (high)} ]\,,
\end{align}
where the first two terms are quadratic in either $M_{\mu\nu}$ or $\delta g_{\mu\nu}^{\rm (high)}$, respectively, and the third term represents the cross terms.
Note that $\delta \Rf{}_{\mu\nu}[M]$ and $\delta \Rff{}_{\mu\nu}[M]$ are the purely low-frequency quantities while $\delta \Rf{}_{\mu\nu}[\delta g^{\rm (high)}]$ and $\delta \Rff{}^{\rm cross}_{\mu\nu}[M\times \delta g^{\rm (high)} ]$ are the purely high-frequency quantities with the inequalities
\begin{align}
\delta \Rf{}_{\mu\nu}[M] &\gg \delta \Rff{}_{\mu\nu}[M] \,, 
\nn
\delta \Rf{}_{\mu\nu}[\delta g^{\rm (high)}]& \gg \delta \Rff{}_{\mu\nu}[M\times \delta g^{\rm (high)} ]\,.
\end{align}
After taking the high/low-frequency projections, we obtain
\begin{align}
\langle R_{\mu\nu} \rangle_{\rm low}&=\RB{}_{\mu\nu}+\delta \Rf{}_{\mu\nu}[M] + \langle \delta \Rff{}_{\mu\nu}[\delta g^{\rm (high)}] \rangle_{\rm low} +\cdots \,, \\
\langle R_{\mu\nu} \rangle_{\rm high} &=\delta \Rf{}_{\mu\nu}[\delta g^{\rm (high)}]+\langle \delta \Rff{}_{\mu\nu}[\delta g^{\rm (high)}] \rangle_{\rm high}+\cdots \,,
\end{align}
 The other quantities are expanded in the similar way and then the equations are decomposed into ones for the low-frequency modes and for the high-frequency modes, respectively.

Up to the linear order the high-frequency mode equations are decomposed into the massless one and the massive one
\begin{align}
\delta \Rf{}_{\mu\nu}[h]-\Lambda_g h_{\mu\nu}
&=\frac{1}{M_{\rm pl}} \left( \delta \Sf{}_{\mu\nu} + \xi_0^2 \delta \cSf{}_{\mu\nu} \right)
\,, 
\label{eom_massless} \\
\delta \Rf{}_{\mu\nu}[\varphi] -\Lambda_g \varphi_{\mu\nu}
+\frac{m_{\rm eff}^2}{4}(2\varphi _{\mu\nu}+\varphi^{\alpha}{}_{\alpha} \gB{}_{\mu\nu})
&=
\frac{1}{M_G} \left( \delta \Sf{}_{\mu\nu} - \alpha^{-1} \xi_0^2 \delta \cSf{}_{\mu\nu} \right) \,,
\label{eom_massive}
\end{align}
where we define
\begin{align}
\delta \Sf{}_{\mu\nu}&:=\left\langle T_{\mu\nu}-\frac{1}{2}g_{\mu\nu} T \right\rangle_{\rm high} \,, \\
\delta \cSf{}_{\mu\nu}&:=\left\langle  \mathcal{T}_{\mu\nu}-\frac{1}{2}f_{\mu\nu}  \mathcal{T} \right\rangle_{\rm high} \,.
\end{align}
Note that $\langle \nablag{}_{\mu} T_{\rm (int)}^{\mu\nu} \rangle_{\rm high}=0$ leads to
\begin{align}
\nablaB{}_{\mu} \varphi^{\mu}{}_{\nu}=\nablaB{}_{\nu} \varphi\,.
\label{const1}
\end{align}
Substituting this into the trace of \eqref{eom_massless}, we find a constraint equation on the trace $\varphi^{\alpha}{}_{\alpha}$
\begin{align}
\left(3m_{\rm eff}^2-2\Lambda_g \right) \varphi^{\alpha}{}_{\alpha}
=
\frac{2}{M_G} \left( \delta \Sf{}_{\alpha\beta} - \alpha^{-1} \xi_0^2 \delta \cSf{}_{\alpha\beta} \right) \gB{}^{\alpha\beta}\,.
\label{const2}
\end{align}

From the low-frequency mode equations, we obtain the Einstein equation for the homothetic background 
\begin{align}
\GB{}^{\mu\nu}+\Lambda_g \gB{}^{\mu\nu} &=\frac{1}{M_{\rm pl}^2} \tau^{\mu\nu}
\,, \label{homothetic_eq}
\end{align}
where $\GB{}^{\mu\nu}$ is the Einstein tensor for $\gB{}_{\mu\nu}$ and
\begin{align}
\tau{}^{\mu\nu}&:=
\TB{}^{\mu\nu}+\xi_0^2 \cTB{}^{\mu\nu}
+\langle T^{\mu\nu}_{\rm gw} \rangle_{\rm low}+ \langle T^{\mu\nu}_G \rangle_{\rm low}
\,, \label{T_tot}
\end{align}
is interpreted as the total energy-momentum tensor including the gravitons as well as the matters. The effective matter energy-momentum tensors for matters $\TB{}_{\mu\nu}$ and $\cTB{}_{\mu\nu}$ are defined by the relations
\begin{align}
\SB{}_{\mu\nu} &:= \left\langle T_{\mu\nu}-\frac{1}{2}g_{\mu\nu} T \right\rangle_{\rm low} 
= \TB{}_{\mu\nu}-\frac{1}{2} \gB{}_{\mu\nu} \TB{}_{\alpha \beta} \gB{}^{\alpha\beta}\,, \\
\cSB{}_{\mu\nu}&:=\left\langle {\mathcal T}_{\mu\nu}-\frac{1}{2}f_{\mu\nu} {\mathcal T} \right\rangle_{\rm low} 
 = \cTB{}_{\mu\nu}-\frac{1}{2} \gB{}_{\mu\nu} \cTB{}_{\alpha \beta} \gB{}^{\alpha\beta}\,,
\end{align}
and the graviton energy-momentum tensors are defined by
\begin{align}
T_{\rm gw}^{\mu\nu}&= -\left( \gB{}^{\mu\alpha} \gB{}^{\nu\beta}-\frac{1}{2}\gB{}^{\mu\nu} \gB{}^{\alpha\beta} \right) 
 \delta \Rff{}_{\alpha\beta}[h] 
\,,
\end{align}
and \eqref{def_TG}.

Finally, we derive the equation for the massive mode $M_{\mu\nu}$ which is given by
\begin{align}
\delta \Rf{}_{\mu\nu}[M]-\Lambda_g M_{\mu\nu}+ \frac{m_{\rm eff}^2}{4}(2M_{\mu\nu}+M^{\alpha}{}_{\alpha} \gB{}_{\mu\nu})
&=\frac{\bar{\kappa}}{M_G} \Delta S_{\mu\nu} \,,
\end{align}
where
\begin{align}
\Delta S_{\mu\nu}:=
 \SB{}_{\mu\nu}-\alpha^{-1}\xi_0^2 \cSB{}_{\mu\nu} 
+ \left\langle -\frac{1}{\kappa_g^2}\delta \Rff{}_{\mu\nu}[\delta g^{\rm (high)}] +\frac{1}{\xi_0^4 \kappa_g^2}\delta \Rff{}_{\mu\nu}[\delta f^{\rm (high)}]
+\Delta \Sff{}_{\mu\nu}^{\rm (int)} \right\rangle_{\rm low}\,,
\end{align}
and
\begin{align}
\Delta \Sff{}_{\mu\nu}^{\rm (int)} =
\frac{m_{\rm eff}^2}{4 \alpha^{1/2}} \left(g_{\mu\nu} h^{\alpha\beta} \varphi_{\alpha\beta}-h_{\mu\nu} \varphi^{\alpha}{}_{\alpha} \right)
-\frac{m_{\rm eff}^2}{16 \alpha}
\Big[ &3(1-\alpha)\gB{}_{\mu\nu} \varphi_{\alpha\beta}\varphi^{\alpha\beta}
+4 \{ (1-\beta_2)\alpha-\beta_2  \}  \varphi_{\mu\nu} \varphi^{\alpha}{}_{\alpha}
\nn
&+2\{ (1+2\beta_2)\alpha-(3-2\beta_2)\alpha \} \varphi_{\mu}{}^{\alpha} \varphi_{\nu\alpha} \Big]
\,.
\end{align}
\end{widetext}
Note that the source term $\Delta S_{\mu\nu}$ is given by the difference between two matter energy-momentum tensors whereas the source term for the massless mode is given by the sum of energy-momentum tensors. The massive mode can give an anti-gravity since the positiveness of the source is not guaranteed even if all energy of the sources are positive definite.


\section{Spacetime deformation as dark matter}
\label{sec_Bianchi}
In this section, we study the axisymmetric Bainchi type I universe and obtain the same conclusion as the main text but from the different picture: we observe the spacetime anisotropy as dark matter in bigravity. In this section, we do not assume either the smallness of the Hubble parameter \eqref{Hubble} or the $Z_2$ symmetry \eqref{symmetric_condition}.

We consider the simplest homogenous but anisotropic universe in bigravity:
\begin{align}
ds_g^2&=-N_g^2 dt^2 +a_g^2[ e^{4\beta_g} dx^2+e^{-2\beta_g}(dy^2+dz^2)]\,, \\
ds_f^2&=-N_f^2 dt^2+a_f^2[ e^{4\beta_f} dx^2+e^{-2\beta_f}(dy^2+dz^2)] \,,
\end{align}
where $\{ N_g,N_f,a_g,a_f,\beta_g,\beta_f \}$ are functions of the time $t$. The Hubble expansion rates and the shears are defined by
\begin{align}
H_g& :=\frac{\dot{a}_g}{a_g N_g} \,, \quad H_f:= \frac{\dot{a}_f}{a_f N_f}\,, \\
\sigma_g& := \frac{\dot{\beta}_g}{N_g}\,, \quad \sigma_f:= \frac{\dot{\beta}_f}{N_f}\,.
\end{align}
Just for simplicity, we consider only the $g$-matter field whose energy-momentum tensor is given by
\begin{align}
T^{\mu}{}_{\nu}={\rm diag}\left[ -\bar{\rho}(t),\bar{p}_{\bot}(t),\bar{p}_{\|}(t),\bar{p}_{\|}(t) \right],
\end{align}
where the pressure is decomposed into the isotropic part $\bar{p}$ and the anisotropic part $\bar{\pi}$:
\begin{align}
\bar{p}&=\frac{1}{3}(\bar{p}_{\bot}+2\bar{p}_{\|})
\,, \\
\bar{\pi}&=\frac{1}{3}(\bar{p}_{\bot}-\bar{p}_{\|})
\,.
\end{align}
The matter is assumed to satisfy the conservation law
\begin{align}
\dot{\bar{\rho}}+3\frac{\dot{a}_g}{a_g}(\bar{\rho}+\bar{p})+6\dot{\beta}_g\bar{\pi}=0\,.
\end{align}

\begin{figure}[htbp]
\centering
\includegraphics[width=6cm,angle=0,clip]{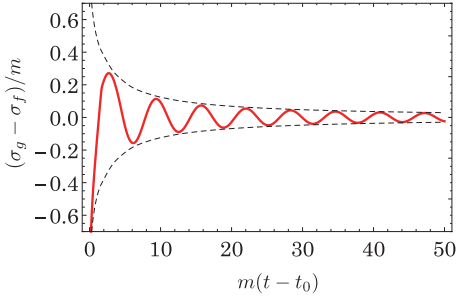}
\caption{The evolution of $\sigma_g-\sigma_f$ in vacuum $\bar{\rho}=\bar{p}=\bar{\pi}=0$ where the coupling constants are chosen as $\kappa_g^2=\kappa_f^2$ and \eqref{coupling} with $b_2=-1$. The initial values are $\beta(t_0)=0.1,\sigma_g(t_0)=-m,\sigma_f(t_0)=0.1m$. The black dashed curves are proportional to $a_g^{-3/2}$.
}
\label{sigma_fig}
\end{figure}

Choosing the gauge $N_g=1$, we find following equations:
the Friedmann equations
\begin{widetext}
\begin{align}
3H_g^2&=\kappa_g^2 \bar{\rho} +3\sigma_g^2+m_g^2 [b_0+b_1(e^{-2\beta }+2e^{\beta}) \xi +b_2(2e^{-\beta}+e^{2\beta}) \xi^2 + b_3 \xi^3 ]
\,, \label{BianchiI_Fri} \\
3H_f^2&=3 \sigma_f^2+m_f^2[b_4 +b_3(2e^{-\beta}+e^{2\beta} ) \xi^{-1} + b_2 (e^{-2\beta}+2 e^{\beta}) \xi^{-2}+b_1 \xi^{-3}]
\,, 
\end{align}
\end{widetext}
the constraint equation
\begin{align}
&H_g \left[ 3b_1+2b_2\xi(2e^{\beta}+e^{-2\beta})+b_3\xi^2 (e^{2\beta}+2e^{-\beta}) \right]
\nn
-&H_f \xi \left[ 3b_3 \xi^2 +2b_2\xi (e^{2\beta}+2e^{-\beta})+b_1(2e^{\beta}+e^{-2\beta}) \right]
\nn
-&2\xi (e^{-\beta}-e^{2\beta})\left[ \sigma_f(b_1e^{-\beta}+b_2 \xi)+\sigma_g(b_2 e^{-\beta}+b_3 \xi) \right]
\nn
&=0 \,,
\label{BianchiI_constraint}
\end{align}
and the equations for anisotropies
\begin{align}
\frac{1}{a_g^3} \frac{d}{dt}\left(a_g^3  \sigma_g\right)+\kappa_g^2 \frac{\partial U_{\beta}}{\partial \beta}&=\kappa_g^2 \bar{\pi} \,,
\label{eq_sigmag}
\\
\frac{1}{a_g^3} \frac{d}{dt}\left( a_f^3 \sigma_f \right) -\kappa_f^2 \frac{\partial U_{\beta}}{\partial \beta}&=0 \,,
\label{eq_sigmaf}
\end{align}
where we define
\begin{align}
\xi & := \frac{a_f}{a_g}\,, \quad
\beta :=\beta_g-\beta_f\,,
\\
m_g^2&:=m^2\frac{\kappa_g^2}{\kappa^2},\quad m_f^2:=m^2\frac{\kappa_f^2}{\kappa^2}\,,
\end{align}
and
\begin{align}
U_{\beta}:=\frac{m^2}{6\kappa^2}
\Big[& \xi \left(2e^{\beta}+e^{-2\beta} \right)(b_1+ b_2N_f)
\nn
&+\xi^2 \left(e^{2\beta}+2e^{-\beta} \right)(b_2+b_3 N_f) \Big]
\,.
\end{align}
Eqs.~\eqref{eq_sigmag} and \eqref{eq_sigmaf} yield
\begin{align}
\frac{1}{a_g^3} \frac{d}{dt}\left[a_g^3 (\kappa_f^2 \sigma_g+\kappa_g^2 \xi^3\sigma_f) \right]=\kappa_g^2 \kappa_f^2 \bar{\pi}\,.
\end{align}
Hence, when the anisotropic stress is ignored $\bar{\pi}=0$, the sum of the shears $\sigma_g$ and $\sigma_f$ decreases as $a_g^{-3}$ which is the same as the standard decaying law of the shear in GR. On the other hand, the difference between them does not decreases as $a_g^{-3}$ due to the ``potential'' $U_{\beta}$. Instead, $\sigma_g-\sigma_f$ (and also $\beta_g-\beta_f$) decrease as $a_g^{-3/2}$ as shown in Fig.~\ref{sigma_fig} and then acts as the ``dark matter'' component of the universe (see also~\cite{Maeda:2013bha}). 

\begin{figure}[htbp]
\centering
\includegraphics[width=6cm,angle=0,clip]{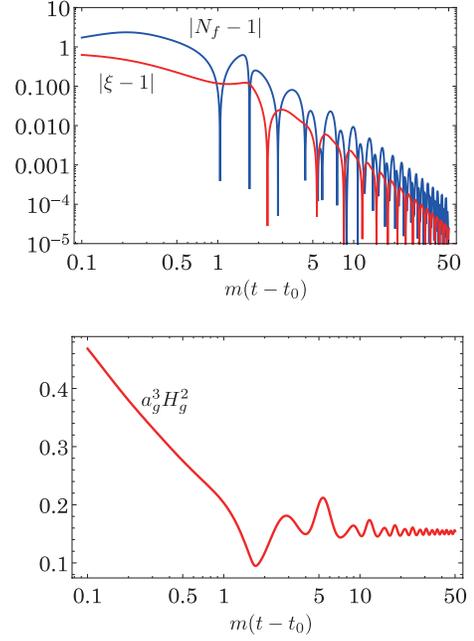}
\caption{The evolutions of $N_f,\xi$ and the rescaled Hubble expansion rate (which is scale free due to the scale factor). We set the same parameters as Fig.~\ref{sigma_fig}.
}
\label{Bianchi1_fig}
\end{figure}

Fig.~\ref{Bianchi1_fig} shows a typical behavior of the solution in vacuum. The top figure represents the spacetimes approach the homothetic solution and the bottom figure shows that $H_g^2$ decreases as $a_g^{-3}$ at the late time where we assume $\Lambda_g|_{\xi_0=1}=0$. Therefore, even if the matter component is not introduced, the same behavior as the dust dominant universe can be obtained.

We then assume $\beta \ll 1$ and consider the regime $m^2 \gg H_g^2, H_f^2$. Note that the smallness of $\beta$ do not suggest that contributions of those to the Friedmann equations are also small because the quantity $m^2\beta^2$ can be comparable to $H_g^2$ and $H_f^2$. We introduce a small quantity $\epsilon:=H_g/m$ and set $\beta =\mathcal{O}(\epsilon)$.
Eq.~\eqref{BianchiI_constraint} gives
\begin{align}
(H_g-H_f \xi)(b_1+2b_2 \xi +b_3 \xi^2) +\mathcal{O}(H_g\epsilon^1)=0\,,
\label{H_g=H_f}
\end{align}
where we have used $\sigma_g,\sigma_f \lesssim H_g,H_f$. 
Since the spacetime can be approximated by the homothetic spacetime in the regime $m^2\gg H_g^2,H_g^2$, the ratio $\xi$ can be expanded around $\xi_0$.
Supposing the normal branch such that
\begin{align}
H_g= \xi H_f+\mathcal{O}(H_g\epsilon) \,,
\end{align}
the Friedmann equations yield
\begin{align}
(3m_{\rm eff}^2-2\Lambda_g) \frac{\xi-\xi_0}{\xi_0}
\approx
- \kappa_g^2 (\bar{\rho}+\bar{\rho}_{\sigma_g})+\xi_0^2 \kappa_f^2 \bar{\rho}_{\sigma_f} \,, \label{delta_xi}
\end{align}
which indicates
$\xi-\xi_0=\mathcal{O}(\epsilon^2) $
where 
\begin{align}
\bar{\rho}_{\sigma_g} & :=\frac{1}{\kappa_g^2}[ 3\sigma_g^2 +3m_g^2 \beta^2 (b_1 \xi_0+b_2 \xi_0^2)]
\,, \\
\bar{\rho}_{\sigma_f}& :=\frac{1}{\kappa_f^2}[ 3\sigma_f^2 +3 m_f^2 \beta^2 (b_2 \xi_0^{-1}+b_3 \xi_0^{-2}) ]
\,.
\end{align}
The deviation of the lapse function is then given by 
\begin{align}
N_f-\xi_0=\frac{1}{H_f}\left( \frac{\dot{\xi}}{\xi}+H_g \right) -\xi_0 = \mathcal{O}(\epsilon^2)\,,
\end{align}
where we notice that, although $\dot{\xi}$ and $H_g$ have quantities of order $\mathcal{O}(\epsilon)$, they are canceled and then $N_f-\xi_0$ is of order $\mathcal{O}(\epsilon^2)$.
Using $m_{\rm eff}^2\gg \Lambda_g$ and including only leading order contributions, the Friedmann equation is expressed by
\begin{align}
3H_g^2 \approx
\Lambda_g+\frac{1}{M_{\rm pl}^2}(\bar{\rho}+\bar{\rho}_{h}+\bar{\rho}_G)
\end{align}
where
\begin{align}
\bar{\rho}_h&:=3\dot{\bar{h}}^2 \,,
\quad
\bar{\rho}_G:=3\left( \dot{\bar{\varphi}}^2+m_{\rm eff}^2 \bar{\varphi}^2\right)\,,
\end{align}
and
\begin{align}
\bar{\varphi}&=\frac{1}{\bar{\kappa}}(\beta_g-\beta_f)\,, \\
\bar{h}&=\frac{\kappa_f}{\xi_0 \kappa_g \bar{\kappa} }\beta_g+\frac{\xi_0 \kappa_g}{\kappa_f \bar{\kappa}} \beta_f\,.
\end{align}
The variables $\bar{\varphi}$ and $\bar{h}$ are the normalized massive mode and the normalized massless mode of the anisotropies, or, following the notion of the main text, they can be interpreted as the ``massive graviton condensate'' and the ``massless graviton condensate'', respectively, which obey
\begin{align}
\ddot{\bar{h}}+3H_g \dot{\bar{h}}&\approx \frac{1}{M_{\rm pl}} \bar{\pi} \,, \\
\ddot{\bar{\varphi}}+3H_g \dot{\bar{\varphi}}+m_{\rm eff}^2 \bar{\varphi}&\approx \frac{1}{M_G} \bar{\pi}\,.
\label{Bianchi_phi}
\end{align}
It is worth noting that although we have assumed the inequalities $\beta \ll 1$ and $m^2\gg H_g^2,H_g^2$ we have not used the high-frequency and the low-frequency projections in the present calculations. 

When we ignore the anisotropic stress, the (averaged) energy densities of $\bar{h}$ and $\bar{\varphi}$ decrease as
\begin{align}
\bar{\rho}_h \propto a_g^{-6}\,,\quad
\langle \bar{\rho}_G \rangle_T \propto a_g^{-3}
\,.
\end{align}
The effect of the homogeneous mode of the massless graviton can be ignored in time.

The averaged differences $\langle \xi-\xi_0 \rangle_T $ and $ \langle N_f-\xi_0 \rangle_T$ correspond to the low-frequency massive mode $M_{\mu\nu}$ which are given by
\begin{align}
\langle  \xi-\xi_0 \rangle_T , \langle N_f-\xi_0 \rangle_T \sim \epsilon^2 \propto m^{-2} \,.
\end{align}
As we expected, $M_{\mu\nu}$ is suppressed by $m^{-2}$ which is just a sub-subleading contribution.

Finally, we discuss the production of the anisotropies. We assume the matter field is composed of radiation and the coherent magnetic field:
\begin{align}
\bar{\rho}_r\propto a_g^{-4}\,, \,\,
\bar{p}_r =\frac{1}{3} \bar{\rho}_r
\,, \,\,
\bar{\pi}_r=0
\,,
\end{align}
and
\begin{align}
\bar{\rho}_B=\frac{\bar{B}^2}{2a_g^4}e^{4\beta_g} 
\,, \,\,
\bar{p}_B=\frac{1}{3}\bar{\rho}_B
\,, \,\,
\bar{\pi}_B=-\frac{2}{3}\bar{\rho}_B
\,,
\end{align}
where $\bar{B}$ is a constant and the strength of the magnetic field is given by $\bar{B}e^{2\beta_g}/a_g^2$. A typical behavior of $\bar{\varphi}$ is shown in Fig.~\ref{production_fig}\footnote{Although the Higuchi instability~\cite{Higuchi:1986py,Higuchi:1989gz,Grisa:2009yy,Fasiello:2012rw,Fasiello:2013woa,Comelli:2012db,Comelli:2014bqa,DeFelice:2014nja} exists in sub-horizon scales in $H_g \gtrsim m$, the unstable modes should not affect the dynamics of the homogeneous mode due to the cosmological Vainshtain mechanism~\cite{Aoki:2015xqa,Mortsell:2015exa,Aoki:2016vcz}. Thus, we may discuss the dynamics of $\bar{\varphi}$ even in $H_g \gtrsim m$.}. The dominant production occurs just after $H_g = m_{\rm eff}$ in which Eq.~\eqref{Bianchi_phi} can be barely used. The produced amplitude is then estimated as
\begin{align}
\bar{\varphi} \sim \frac{\bar{\pi}}{M_G m_{\rm eff}^2}
\,.
\end{align}
In the main text, since we have assumed $M_G=M_{\rm pl}$ and used the normalization $m_{\rm eff}=m$, we obtain \eqref{produced_phi}.


\section{General perturbations}
\label{sec_general_pert}
In this section, we summarize the calculations for the general perturbations around the homogeneous solutions \eqref{FLRW_ansatz} and \eqref{coherent_phi}. We use the specific choice of the coupling constants $\kappa_g^2=\kappa_f^2$ and \eqref{coupling} to simplify the calculations.

\subsection{Harmonics expansion}
The homogeneity of the background solution leads to that all variables can be transformed into the momentum space. Furthermore, due to the rotational symmetry in the $y$-$z$ space, the perturbations are decomposed into the even parity perturbations and the odd parity perturbations
and there is no interaction between the even and the odd parity perturbations at linear order.
The even and odd parity perturbations are parametrized as
\begin{widetext}
\begin{align} 
\delta \varphi_{\mu\nu}^{\rm (even)} &=
\begin{pmatrix}
-2 \phi Y_S  & -a B Y_x  & -a C Y_a \\
*& 2a^2 (\psi + 2\delta \varphi )Y_S & 2a^2 DY_{xa} \\
*&* & a^2 [2(\psi-\delta \varphi) \delta_{ab} Y_S +2 EY_{ab}]
\end{pmatrix}
\,, \\
\delta \varphi^{\rm (odd)}_{\mu\nu} &=
\begin{pmatrix}
 0 & 0 & -a   \mathcal{B} \mathcal{Y}_a \\
 * & 0 & a^2  \mathcal{D} \mathcal{Y}_{xa}\\
 * & * & 2a^2 \mathcal{E} \mathcal{Y}_{ab}
\end{pmatrix}
\,,
\end{align}
\end{widetext}
where
\begin{align}
Y_x&=-k^{-1} \partial_x Y_S \,, \nn 
Y_a&=-k^{-1} \partial_a Y_S
\,, \nn
Y_{xa}&=k^{-2}\partial_x \partial_a Y_S \,, \nn
Y_{ab}&=k^{-2} \left( \partial_a \partial_b -\frac{1}{2}\delta_{ab} \partial^c \partial_c \right) Y_S
\,, \label{even_harmonics}
\end{align}
and
\begin{align}
\mathcal{Y}_a&=\frac{1}{k}\epsilon_{a}{}^{b} \partial_b Y_S
\,, \nn
\mathcal{Y}_{xa}&=-\frac{1}{k} \partial_x \mathcal{Y}_a=-\frac{1}{k^2} \partial_x \epsilon_{b}{}^c \partial_c Y_S
\,, \nn
\mathcal{Y}_{ab}&=-\frac{1}{k}\partial_{(a}\mathcal{Y}_{b)}=-\frac{1}{k^2}\partial_{(a} \epsilon_{b)}{}^c \partial_c Y_S
\,,
\end{align}
with $Y_S=e^{ikx}$.

For our study, it is useful to define the harmonics associated with the three dimensional Euclidean space. We define a three dimensional vector and a tensor as
\begin{align}
Y_{Vi}&=k^{-2}\left( k_{\|}^2 Y_x, -k_x^2Y_a \right)
\,,
\\
Y_{Tij}&=k^{-4}
\begin{pmatrix}
k_{\|}^4 Y_S & k^2k_{\|}^2 Y_{xa} \\
* & -\frac{k_{\|}^4}{2}Y_S \delta_{ab}-(k^4+k_x^2k^2)Y_{ab}
\end{pmatrix}
\,,
\end{align}
and
\begin{align}
\mathcal{Y}_V{}_i&=(0,  \mathcal{Y}_a)
\,, \\
\mathcal{Y}_T{}_{ij}&=k^{-2}
\begin{pmatrix}
0 & k_{\|}^2\mathcal{Y}_{xa} \\
* & -2k_x^2 \mathcal{Y}_{ab}
\end{pmatrix}
\,,
\end{align}
which satisfy \eqref{SVT_harmonics1} and \eqref{SVT_harmonics2}.
Using the three dimensional harmonics, any three dimensional vector (or tensor) is uniquely decomposed into the scalar and the vector (and the tensor) quantities. For example, the even parity perturbations and the odd parity perturbations of the graviton energy-momentum tensor are given by \eqref{TG_even} and \eqref{TG_odd}, respectively.

\subsection{Perturbations for massless mode}
The low-frequency mode $\gB{}_{\mu\nu}$ has the gauge symmetry by which we can obtain \eqref{delg_even} and \eqref{delg_odd}.

The general form of the even parity metric perturbations is given by
\begin{align}
\delta g{}_{\mu\nu}=
\begin{pmatrix}
-2\Phi Y_S & -a B_i \\
* & a^2\left[ 2\Psi Y_S \delta_{ij}+H_{ij} \right]
\end{pmatrix}
\,,
\end{align}
with
\begin{align}
B{}_i=(B_g Y_x , C_g Y_a)\,,
\end{align}
and
\begin{align}
H{}_{ij}=
\begin{pmatrix}
4hY_S & 2D_g  Y_{xb} \\
* & -2hY_S\delta_{ab} +2E_g Y_{ab}
\end{pmatrix}
\,,
\end{align}
where we note
\begin{align}
H^i{}_i=0\,,
\end{align}
and
\begin{align}
\partial_i B^i&= (k_x^2 B_g+k_{\|}^2 C_g ) k^{-1} Y_S
\,, \\
\partial_i H^{ix}&= \left( 2 k_{\|}^2 D_g- 4 k^2 h_g \right) k^{-1} Y_x
\,, \nn
\partial_i H^{ia}&=\left( 2k_x^2 D_g +k_{\|}^2 E_g +2k^2 h_g\right) k^{-1} Y^a
\,.
\end{align}

Under the gauge transformation
\begin{align}
x^{\mu}\rightarrow x^{\mu}+\xi^{\mu}_{\rm (even)}
\,,
\end{align}
with
\begin{align}
\xi^{\mu}=(\xi^0 Y_S, \xi^1 Y_x , \xi^2 Y^a )
\,,
\end{align}
the perturbations transform as
\begin{align}
\Phi_g &\rightarrow \Phi_g+\dot{\xi}^0 \,, \nn
\Psi_g &\rightarrow \Psi_g +H\xi^0+\frac{1}{3}k \xi_L \,, \nn
h_g &\rightarrow h_g+\frac{k_x^2}{3k} \xi^1-\frac{k_{\|}^2}{6k} \xi^2 \,, \nn
B_g &\rightarrow B_g-\frac{k}{a}\xi^0-a\dot{\xi}^1 \,, \nn
C_g &\rightarrow C_g-\frac{k}{a}\xi^0-a\dot{\xi}^2 \,, \nn
D_g &\rightarrow D_g-\frac{k}{2}(\xi^1+\xi^2) \,, \nn
E_g &\rightarrow E_g-k\xi^2\,,
\end{align}
and then
\begin{align}
\partial_i B^i  &\rightarrow \partial_i B^i-\left( \frac{k^2}{a} \xi^0 +ak \dot{\xi}_L \right) Y_S
\,, \nn
\partial_i H^{ix} &\rightarrow \partial_i H^{ix}- \left(  \xi^1+\frac{1}{3} \xi_L \right) k^2 Y_x
\,, \nn
\partial_i H^{ia} &\rightarrow \partial_i H^{ia}-\left( \xi^2+\frac{1}{3}\xi_L \right) k^2 Y^a
\,,
\end{align}
where
\begin{align}
k^2 \xi_L=k_x^2 \xi^1+k_{\|}^2 \xi^2
\,.
\end{align}
Therefore one can fix the gauge so that
\begin{align}
\partial_i B^i=0\,,\quad
\partial_i H^{ij}=0
\,,
\end{align}
in which $B^i$ and $H^{ij}$ are given by
\begin{align}
B^i=B_V Y_V^i\,, \quad
H^{ij}=2H_T Y^{ij}_T\,,
\end{align}
where $B_V$ and $H_T$ are functions of time.

The odd parity metric perturbations are
\begin{align}
\delta g_{\mu\nu}^{\rm (odd)}
=
\begin{pmatrix}
0 & -a \mathcal{B}_ {i} \\
* & a^2 \mathcal{H}_ {ij} 
\end{pmatrix}
\,,
\end{align}
with
\begin{align}
\mathcal{B}_{i}=(0,  \mathcal{B}_g \mathcal{Y}_a)\,,
\end{align}
and
\begin{align}
\mathcal{H}_{ij}=
\begin{pmatrix}
0 &  \mathcal{D}_g \mathcal{Y}_{xb}
\\
 * & 2 \mathcal{E}_g \mathcal{Y}_{ab}
\end{pmatrix}
\,,
\end{align}
where we notice
\begin{align}
\partial_i \mathcal{B}^i&=0 \,, \\
\partial_i \mathcal{H}^{ix}&=0\,, \\
\partial_i \mathcal{H}^{ia}&= \left( k_x^2 \mathcal{D}_g+k_{\|}^2 \mathcal{E}_g \right) k^{-1} \mathcal{Y}^a
\,,\\
\mathcal{H}^i{}_i&=0\,.
\end{align}

Under the gauge transformation
\begin{align}
x^{\mu}\rightarrow x^{\mu}+\xi^{\mu}_{\rm (odd)}\,,
\end{align}
with
\begin{align}
\xi^{\mu}_{\rm (odd)}=(0, 0 ,  \xi_{\rm odd} \mathcal{Y}^a)\,,
\end{align}
the metric perturbations transform as
\begin{align}
\mathcal{B}_g&\rightarrow \mathcal{B}_g-a \dot{\xi}_{\rm (odd)}
\,, \\
\mathcal{D}_g&\rightarrow \mathcal{D}_g- k \xi_{\rm (odd)}
\,, \\
\mathcal{E}_g&\rightarrow \mathcal{E}_g- k \xi_{\rm (odd)}
\,,
\end{align}
and then
\begin{align}
\partial_i \mathcal{H}^{ia}\rightarrow \partial_i \mathcal{H}^{ia}
- k^2 \xi_{\rm (odd)}  \mathcal{Y}^a
\,.
\end{align} 
One can choose the gauge
\begin{align}
\partial_i \mathcal{H}^{ia}&
=0\,,
\end{align}
in which $\mathcal{B}_g^i$ and $\mathcal{H}_g^{ij}$ corresponds to the vector perturbation and the tensor perturbation, respectively:
\begin{align}
\mathcal{B}^i=\mathcal{B}_V \mathcal{Y}^i_V
\,,\quad
\mathcal{H}_{ij}=2\mathcal{H}_{T} \mathcal{Y}_T^{ij}\,.
\end{align}

\subsection{Adiabatic expansion}

In addition to the linearization of the inhomogeneities, we shall take the adiabatic expansion in terms of $m^{-1}$. To verify the linearization the amplitudes of the inhomogeneities are of $O(m^n)$ with $n\leq 0$. 

The equation \eqref{wave_eq_order_m} yields that, up to the subleading order, the massive graviton can be expressed as
\begin{align}
\delta \varphi_{\mu\nu}=\delta \varphi_1{}_{\mu\nu} \cos[mt]+\delta \varphi_2{}_{\mu\nu} \sin[mt]
\,,
\end{align}
where $\delta \varphi_1{}_{\mu\nu}$ and $\delta \varphi_2{}_{\mu\nu}$ are slowly varying function in time\footnote{If we do not assume the $Z_2$ symmetry for the self-interactions, the frequency at the subleading order can differ from $m/2 \pi$.}. We use the suffixes $1$ and $2$ to represent the slowly varying functions in front of $\cos[m t]$ and $\sin[m t]$ (see \eqref{slowly_functions}). The orders of $\delta \varphi_1{}_{\mu\nu}$ and $\delta \varphi_2{}_{\mu\nu}$ are determined to be consistent with the equations.

\subsection{Odd parity}
We first study the odd parity perturbations in which \eqref{traceless_order_m} is trivially satisfied. We discuss the large scales \eqref{large_scale} and the small scales \eqref{small_scale} in order.

\subsubsection{Large scales}
The consistency of the equations yield
\begin{align}
\mathcal{B}_{1,2}=\mathcal{O}(m^{-1})\,,
\end{align}
and other variables are of order $\mathcal{O}(m^0)$.

Form the equations of motion \eqref{wave_eq_order_m} and \eqref{transverse_order_m} we obtain the constraint equations
\begin{align}
\frac{mk}{a}\mathcal{B}_2+\frac{k_x^2}{a^2} \mathcal{D}_1+\frac{k_{\|}^2}{a^2} \mathcal{E}_1&=0
\,, \\
\frac{mk}{a}\left( \mathcal{B}_1+2\bar{\varphi}_1 \mathcal{B}_V \right)-\frac{k_x^2}{a^2} \mathcal{D}_2-\frac{k_{\|}^2}{a^2} \mathcal{E}_2&=0
\,,
\end{align}
and the dynamical equations
\begin{align}
\dot{\mathcal{D}}_{1,2}+\frac{3}{2}H\mathcal{D}_{1,2}&=0
\,, \\
\dot{\mathcal{E}}_{1,2}+\frac{3}{2}H\mathcal{E}_{1,2}&=0
\,,
\end{align}
which leads to
\begin{align}
\mathcal{D}_{1,2},\mathcal{E}_{1,2} \propto a^{-3/2}
\,.
\end{align}

Then, we obtain the velocity and the pressure of the massive graviton condensate as
\begin{align}
v_{\rm (odd)}=\mathcal{B}_V-\frac{1}{2m \bar{\varphi}_1} \frac{k_x^2}{ak}\mathcal{D}_2
\end{align}
and
\begin{align}
\pi_V^{\rm (odd)}=\pi_T^{\rm (odd)}=\mathcal{O}(m^{-1})
\,,
\end{align}
which indicates that the massive graviton can be interpreted as the pressureless ideal fluid.

\subsubsection{Small scales}
From the consistency we find
\begin{align}
\mathcal{B}_{1,2}=\mathcal{O}(m^{-1/2})
\,, \quad
\mathcal{B}_V=\mathcal{O}(m^{-1/2})\,,
\end{align}
and others are of order $\mathcal{O}(m^0)$.

The constraint equations are given by
\begin{align}
\frac{mk}{a}\mathcal{B}_{1,2}\mp \frac{k_x^2}{a^2}\mathcal{D}_{2,1}\mp \frac{k_{\|}^2}{a^2}\mathcal{E}_{2,1}=0\,,
\end{align}
while the dynamical equation are
\begin{align}
\dot{\mathcal{D}}_{1,2}+\frac{3}{2}H\mathcal{D}_{1,2}\mp \frac{k^2}{2ma^2}\mathcal{D}_{2,1}&=0
\,, \label{eq_D}\\
\dot{\mathcal{E}}_{1,2}+\frac{3}{2}H\mathcal{E}_{1,2}\mp \frac{k^2}{2ma^2}\mathcal{E}_{2,1}&=0
\,. \label{eq_E}
\end{align}
The solutions are
\begin{align}
\mathcal{D}_{1,2},\mathcal{E}_{1,2} \propto a^{-3/2} \exp \left[ \pm i \frac{3k^2}{2ma^2}t \right]
\,,
\end{align}
when $a\propto t^{2/3}$ (the dust dominant universe).

We then obtain the velocity
\begin{align}
v_{\rm (odd)}=-\frac{1}{2m\bar{\varphi}_1} \frac{k_x^2}{ak} \mathcal{D}_2 \,,
\end{align}
and the anisotropic pressures
\begin{align}
\pi^{\rm (odd)}_V&=-\frac{3k_x^2}{2a^2}\bar{\varphi}_1 \mathcal{D}_1 
\,, \\
\pi^{\rm (odd)}_T&=-\frac{k^2}{4a^2}\bar{\varphi}_1 (\mathcal{D}_1+2\mathcal{E}_1)
\,.
\end{align}
Since $\mathcal{D}_{1,2}$ and $\mathcal{E}_{1,2}$ are independent, we can choose $\mathcal{D}_{1,2}$ and $\mathcal{D}_{1,2}+2\mathcal{E}_{1,2}$ as new variables which obey the same equation as \eqref{eq_D} and \eqref{eq_E}. The vector quantities $v_{\rm (odd)}$ and $\pi_V^{\rm (odd)}$ evolve independently from the tensor quantity $\pi^{\rm (odd)}_T$. As a result, the vector and the tensor modes are decoupled.

\subsection{Even parity}

\subsubsection{Large scales}
In the large scales \eqref{large_scale}, we obtain
\begin{align}
\phi_{1,2},\psi_2&=\mathcal{O}(m^{-2})
\,, \nn
B_{1,2},\psi_{1},\delta \varphi_1 &=\mathcal{O}(m^{-1})\,,
\end{align}
and others are of order $\mathcal{O}(m^0)$. Since the quantities of order $\mathcal{O}(m^{-2})$ are sub-subleading order contributions, $\phi_{1,2}$ and $\psi_2$ are irrelevant to the dynamics.

From \eqref{transverse_order_m} and \eqref{traceless_order_m} we find five constraint equations 
\begin{align}
2k_{\|}^4 H_T \bar{\varphi}_1-k^4 \psi_1&=0 
\,, \\
\frac{mk}{a}B_2+\frac{2k_{\|}^2}{a^2}D_1&=0
\,,\\
\frac{mk}{a}C_2+\frac{2k_x^2}{a^2}D_1+\frac{k_{\|}^2}{a^2}E_1&=0
\,, \\
\frac{mk}{a}B_1-\frac{4mk_{\|}^2}{ak}B_T \bar{\varphi}_1-\frac{2k_{\|}^2}{a^2}D_2+4\frac{k^2}{a^2}\delta \varphi_2 &=0
\,, \\
-\frac{mk}{a}C_1+\frac{2mk_x^2}{ak}B_T \bar{\varphi}_1+\frac{2k_x^2}{k^2}D_2&
\nn
+\frac{k_{\|}^2}{a^2}E_2+\frac{2k^2}{a^2}\delta \varphi_2&=0
\,,
\end{align}
which determines $\{ \psi_1,B_{1,2}, C_{1,2}\}$ in terms of other variables.
The non-trivial components \eqref{wave_eq_order_m} yield
\begin{align}
\delta \dot{\varphi}_2+\frac{3}{2}H\varphi_2+m\Phi \bar{\varphi}_1=0\,,
\end{align}
and
\begin{align}
\dot{D}_{1,2}+\frac{3}{2}HD_{1,2}=0
\,, \\
\dot{E}_{1,2}+\frac{3}{2}HE_{1,2}=0\,.
\end{align}
We cannot find other equations within our accuracy.

We note that the amplitudes of $D_{1,2}$ and $E_{1,2}$ always decreases since the gravitational potential $\Phi$ does not affect the dynamics of $D_{1,2}$ and $E_{1,2}$. The Jeans instability does not lead to the growth of $D_{1,2}$ and $E_{1,2}$. 

After taking the oscillation average, we find that the pressures are zero. The energy density and $v_S$ are given by \eqref{rhoG_large} and \eqref{vS_large}, respectively. The vector mode of the velocity is 
\begin{align}
v_V&=B_V+\frac{k}{am\bar{\varphi}_1} D_2
\,.
\end{align}
The massive graviton condensate is a form of a pressureless perfect fluid. Clearly from the definitions of $\delta \rho_G, v_S$ and $v_V$, the vector mode is decoupled from other modes.

As already mentioned in the main text, the evolution of $\delta \varphi_1$ is not determined. However, the dynamics of $\delta \rho_G$ is determined by the conservation law of the averaged graviton energy-momentum tensor.

\subsubsection{Small scales}
Finally, we study the even parity perturbations in the small scales \eqref{small_scale}. The consistency leads to
\begin{align}
B_V&=\mathcal{O}(m^{-1/2})\,, 
\\
B_{1,2},C_{1,2}&=\mathcal{O}(m^{-1/2})
\,, \nn
\phi_{1,2},\psi_{1,2}&=\mathcal{O}(m^{-1})\,,
\end{align}
and others are of order $\mathcal{O}(m^0)$.

We obtain the eight constraint equations in terms of the eight variables $\{ \psi_{1,2}, \phi_{1,2}, B_{1,2}, C_{1,2} \}$ as follows:
\begin{align}
\phi_1+3\psi_1-6\frac{k_{\|}^4}{k^4}H_T \bar{\varphi}_1&=0
\,, \\
\phi_2+3\psi_2&=0
\,, \\
\pm \frac{2mk}{a}\phi_{1,2}+\frac{k_x^2}{a^2}B_{2,1}+\frac{k_{\|}^2}{a^2}C_{2,1}&=0
\,, \\
\pm \frac{mk}{a} B_{1,2}+2\frac{k_{\|}^2}{a^2}D_{2,1}-4\frac{k^2}{a^2}\delta \varphi_{2,1}&=0
\,, \\
\pm \frac{mk}{a}C_{2,1}+2\frac{k_x^2}{a^2}D_{1,2}+\frac{k_{\|}^2}{a^2}E_{1,2}+2\frac{k^2}{a^2}\delta \varphi_{1,2}&=0
\,.
\end{align}
Eq.~\eqref{wave_eq_order_m} gives six equations
\begin{align}
\delta \dot{\varphi}_1+\frac{3}{2}H \delta \varphi_1 -\frac{k^2}{2ma^2}\delta \varphi_2 &=0
\,, \\
\delta \dot{\varphi}_2+\frac{3}{2}H \delta \varphi_2 +\frac{k^2}{2ma^2}\delta \varphi_1+m \Phi \varphi_1&=0
\,, \\
\dot{D}_{1,2}+\frac{3}{2}H D_{1,2} \mp \frac{k^2}{2ma^2}D_{2,1}&=0
\,, \\
\dot{E}_{1,2}+\frac{3}{2}H E_{1,2} \mp \frac{k^2}{2ma^2}E_{2,1}&=0
\,,
\end{align}
which determine the dynamics of $\{ \delta \varphi_{1,2}, D_{1,2},E_{1,2} \}$.

The energy density and the velocity are expressed as \eqref{rhoG_small}, \eqref{vS_small} and
\begin{align}
v_V&=\frac{k}{am\bar{\varphi}_1}D_2
\,.
\end{align}
\begin{widetext}
The pressures are not zero and given by
\begin{align}
\delta p_G&=-\frac{\bar{\varphi}_1}{6a^2k^2}
\left[ k^2 \left( 11k_x^2 +5 k_{\|}^2 \right)\delta \varphi_1
-2k_x^2k_{\|}^2D_1+k_{\|}^4 E_1 \right]\,,
\\
\pi_S&=-\frac{\bar{\varphi}_1}{4a^2k^2}\left[ 2k^2 \left(10k_x^2+7k_{\|}^2 \right) \delta \varphi_1
-2k_x^2k_{\|}^2 D_1+k_{\|}^4 E_1 \right]
\,,
\\ 
\pi_V&=3\frac{k^2}{a^2}\bar{\varphi}_1D_1
\,, \\
\pi_T&=-\frac{\bar{\varphi}_1}{4a^2}
\left[ 6 k^2 \delta \varphi_1+
2k_x^2D_1+
\left(2k_x^2+k_{\|}^2\right)E_1
\right]\,,
\end{align}
Since the vector type perturbations are determined by $D_1$ and $D_2$ only, the vector perturbations are decoupled from other perturbations. However, the scalar type perturbations and the tensor type perturbations are affected by $\delta \varphi_{1}, D_1$ and $E_1$ and then they are not decoupled.
\end{widetext}


\bibliography{ref}
\bibliographystyle{JHEP}

\end{document}